\def\rf#1{(\ref{#1})}
\def\tph{\widetilde{\eta}}
\def\tu{\widetilde{u}}
\def\tf{\widetilde{f}}
\def\txi{\widetilde{\xi}}
\def\wf{\widehat f}
\def\bom{\mbox{\boldmath$\omega$\unboldmath}}
\def\la{\langle}
\def\ra{\rangle}
\documentclass[3p]{elsarticle}
\usepackage{epsfig}
\biboptions{square,comma,sort&compress}
\begin{document}
\title{The Monge--Amp\`ere equation: various forms and numerical solution}

\author[mitpan,oca]{V.~Zheligovsky}
\ead{vlad@mitp.ru}
\cortext[cor]{Corresponding author}

\author[mitpan,oca]{O.~Podvigina}
\ead{olgap@mitp.ru}

\author[oca]{U.~Frisch}
\ead{uriel@oca.eu}

\address[mitpan]{International Institute of Earthquake Prediction Theory and
Mathematical Geophysics,\\84/32 Profsoyuznaya St, 117997 Moscow, Russian Federation}

\address[oca]{UNS, CNRS, Laboratoire Cassiop\'ee, Observatoire de la C\^ote d'Azur\\
BP 4229, 06304 Nice Cedex 4, France}

\begin{abstract}
We present three novel forms of the Monge--Amp\`ere equation, which is
used, e.g., in image processing and in reconstruction of mass
transportation in the primordial Universe. The central role in this
paper is played by our Fourier integral form, for which we establish
positivity and sharp bound properties of the kernels. This is the
basis for the development of a new method for solving numerically the
space-periodic Monge--Amp\`ere problem in an odd-dimensional
space. Convergence is illustrated for a test problem of cosmological
type, in which a Gaussian distribution of matter is assumed in each
localised object, and the right-hand side of the Monge--Amp\`ere
equation is a sum of such distributions.
\end{abstract}


\maketitle

\section{Introduction}

The Monge--Amp\`ere equation (MAE)
\begin{equation}
\det\|u_{x_ix_j}\|=f({\bf x})
\label{MAeq}\end{equation}
is encountered in many areas of numerical analysis and physics, ranging from
image processing \cite{HTK,HZTA,HGS} to cosmology \cite{Fr02,Br03,Mo08}.
Here the subscripts $x_i$ denote derivatives in
the respective spatial variables; $\|u_{x_i,x_j}\|$ is the Hessian of $u$, i.e.,
the matrix comprised of second derivatives of $u$. Existence and regularity
of its solutions was considered in \cite{GiTr83,Bak94,Caf}. Various strategies
were proposed for its numerical solution. A provably convergent method
for solving the Dirichlet problem for the two-dimensional MAE in bounded convex
domains was presented in \cite{OP88}. It is directly linked to the geometric
interpretation of the equation that had given an opportunity to demonstrate
existence of weak solutions \cite{P75}. However, the actual numerical examples
in \cite{OP88} have just up to 25 grid points, which is clearly insufficient
to establish the practical feasibility of the method.\footnote{More recently,
a similar algorithm was used \cite{beam} to design a beam--shaping reflector;
computations on a $55\times 55$ grid required about 15 minutes of a Pentium 4
processor. This problem of geometric optics can be recast as
the Monge--Kantorovich mass transfer problem, reducing to a PDE of the MAE
type, see \cite{GlOl}.} The MAE can be
recast as a minimisation problem; application of algorithms for saddle-point
optimisation to a two-dimensional MAE was considered in \cite{BB00,DG03}, and
least-squares minimisation was advocated in \cite{DG04}. Efficient methods
for solution of the discrete optimal transportation problem were developed for
application to cosmological problems \cite{Fr02,Br03,Mo08}. It may be perceived
that discrete methods correspond well to the physics of mass transportation
in the Universe --- after all, galaxies, as observed by astronomers, are clearly
well-localised, discrete objects! This argument, however, becomes less persuading,
when one recalls that visible matter constitutes only a very small fraction of
different, ``dark'' kinds of matter, whose density distribution is supposed
to be continuous (as opposed to a nearly discrete distribution in clusters
of the visible matter).

In another group of methods the MAE is treated as a generic nonlinear partial
differential equation. Application of Galerkin's and finite element methods to
a regularised MAE was considered in \cite{Feng07a,Feng07b}. A pseudospectral
Newton's algorithm was reported to perform well for a MAE in $R^2$ with
a smooth r.h.s.~\cite{LR05}. However, when we implemented this algorithm for
a three-dimensional MAE, we found that it failed to converge, unless the
r.h.s.~was a smooth slowly varying function. This is consistent with the observation
\cite{OP88} that Newton's method, applied to equations equivalent to the MAE,
is useful to improve an approximate solution only if the approximation is
accurate enough, and easily fails otherwise. Two methods for numerical solution
of a Dirichlet problem for an elliptic MAE in a two-dimensional convex region
$\Omega$ are examined in \cite{BFO09}. The first one employs a finite-difference
discretisation of the equation; it is advocated for application to the MAE with
a possibly singular solution. The second one is an iterative method for the MAE
in the form of a fixed point problem
$$u=\nabla^{-2}\sqrt{u^2_{x_1x_1}+u^2_{x_2x_2}+2u^2_{x_1x_2}+2f}$$
(here $\nabla^{-2}$ denotes the inverse Laplacian); it is claimed to perform
better, when the solution is regular (i.e., belongs to the Sobolev space
$W^2_2(\Omega)$\,).

A so-called ``inexact'' iterative Newton--Krylov solver with preconditioning
was applied for two- \cite{Del08} and three-dimensional \cite{Finn} grid
generation with the properties of equidistribution and minimum distortion.
The nature of the problem solved in \cite{Del08,Finn} required only a modest
accuracy of solutions, discrepancies the order of $10^{-3}-10^{-4}$ being
acceptable. Examples of computations of such moderately accurate solutions
illustrating the scalability of the algorithm with respect to the employed
spatial resolution were provided {\it ibid.}

In Sections 2-4 we derive three alternative forms of \rf{MAeq}:
the ``second-order divergence'' form, the Fourier integral form and
the ``convolution'' form, which, to the best of our knowledge, were never
presented in the literature before. The second form suggests
two related methods for computation of space-periodic solutions to \rf{MAeq}.
The methods and results of their test applications are presented in Section 6;
the Monge-Amp\`ere problem of the cosmological type, which we use to test our
algorithms, is presented in Section 5, after a short statement of the
cosmological reconstruction problem. In the Concluding remarks we briefly
discuss some open questions.

We note that the MAE's of various kinds are considered in the literature.
For the MAE
arising in the mass transportation problem, typically the r.h.s.~of \rf{MAeq}
depends on the gradient of the unknown function (this form of the equation can
be found in the original memoir by Amp\`ere \cite{Amp}). In the theory of
PDE's, usually one considers the MAE with a known r.h.s.~(see
\cite{Caf,Gut}). The MAE of this kind arises in the reconstruction of the early
Universe (see discussion at the beginning of Section 5). Since this
cosmological problem is the main original motivation for our work, we restrict
our discussion to the case of the ``mathematician's'' MAE. However,
a straightforward reformulation of our algorithms makes them applicable
for the case of the r.h.s.~depending on the unknown function.

\section{The ``second-order divergence'' form of the MAE}

The divergence form of the MAE \rf{MAeq}, in which the l.h.s.~of the equation
is represented as a sum of the {\it first} derivatives of certain quantities, is well
known (see, e.g., \cite{Feng07a}). In this section we derive a representation of
the l.h.s.~of the MAE as a sum of {\it second} derivatives, which we call
the second-order divergence form of the MAE.

Consider a Fourier integral solution
$$u=\int_{R^N}\tu(\bom)e^{i\bom\cdot\bf x}\,d\bom$$
to the MAE in $R^N$. Substituting the integral into \rf{MAeq}
and using the identity for $N\times N$ matrices
$$\det\|a_{ij}\|={1\over N!}\,\sum_{i_1,...,i_N,j_1,...,j_N}
\varepsilon_{i_1...i_N}\varepsilon_{j_1...j_N}\prod_{n=1}^Na_{i_nj_n}$$
where each of the indices $i_1,...,i_N,j_1,...,j_N$ takes the values
$1,...,N$ and $\varepsilon_{p_1...p_N}$ denotes the unit antisymmetric tensor
of rank $N$, we find
$$\det\|u_{x_ix_j}\|={(-1)^N\over N!}\,\sum_{i_1,...,i_N,j_1,...,j_N}
\varepsilon_{i_1...i_N}\varepsilon_{j_1...j_N}\prod_{n=1}^N\int_{R^N}
\tu(\bom)\omega_{i_n}\omega_{j_n}e^{i\bom\cdot\bf x}\,d\bom$$
$$={(-1)^N\over N!}\,\int_{R^N}...\int_{R^N}\left(\,\sum_{i_1,...,i_N}
\varepsilon_{i_1...i_N}\prod_{n=1}^N\omega^n_{i_n}\right)
\left(\,\sum_{j_1,...,j_N}\varepsilon_{j_1...j_N}\prod_{n=1}^N\omega^n_{j_n}
\right)\left(\,\prod_{n=1}^N\tu(\bom^n)\right)$$
$$\times\exp\left(i\sum_{n=1}^N\bom^n\cdot{\bf x}\right)\,d\bom^1...\,d\bom^N$$
$$={(-1)^N\over N!}\,\int_{R^N}...\int_{R^N}{\det}^2
\left\|\bom^1,...,\bom^{N-1},\bom-\sum_{n=1}^{N-1}\bom^n\right\|$$
\begin{equation}
\times\left(\,\prod_{n=1}^{N-1}\tu(\bom^n)\right)
\tu\!\left(\bom-\sum_{n=1}^{N-1}\bom^n\right)
e^{i\bom\cdot\bf x}\,d\bom^1...\,d\bom^{N-1}d\bom
\label{detone}\end{equation}
\begin{equation}
\!={(-1)^N\over N!}\!\int_{R^N}\!...\!\int_{R^N}\!{\det}^2
\!\left\|\bom^1\!\!,...,\bom^{N-1}\!\!,\bom\right\|
\left(\,\prod_{n=1}^{N-1}\tu(\bom^n)\!\right)
\tu\!\left(\!\bom-\!\!\sum_{n=1}^{N-1}\bom^n\!\right)\!
e^{i\bom\cdot\bf x}d\bom^1...\,d\bom^{N-1}d\bom.
\label{dettwo}\end{equation}
Here $\left\|\bom^1,...,\bom^N\right\|$ denotes a $N\times N$ matrix comprised
of $N$ columnar vectors $\bom^1,...,\bom^N$. The first factor in the integrand
of \rf{dettwo} is a quadratic function of $\bom$, hinting that the l.h.s.~of
\rf{MAeq} can be transformed into a sum of second derivatives. Indeed, ``reverse
engineering'' of \rf{dettwo} reveals an alternative,
{\it ``second-order divergence'' form} of \rf{MAeq} in $R^N$:
\begin{equation}
{1\over N!}\,\sum_{i_1,...,i_N,j_1,...,j_N}
\varepsilon_{i_1...\,i_N}\varepsilon_{j_1...\,j_N}
\left(u_{x_{i_1}x_{j_1}}...\,u_{x_{i_{N-1}}x_{j_{N-1}}}u\right)_{x_{i_N}x_{j_N}}=f.
\label{diver}\end{equation}
Equivalence of \rf{MAeq} and \rf{diver} is now easily established directly:
Using the standard rule for differentiation of the products in the l.h.s.~of
\rf{diver}, we render it as a sum of products of derivatives of $u$.
Third-order derivatives enter such a product only in pairs of the form
$u_{x_{i_p}x_{i_N}x_{j_p}}u_{x_{i_q}x_{j_N}x_{i_q}}$, and hence all products
involving third-order derivatives cancel out due to antisymmetry of the tensors
$\varepsilon_{i_1...\,i_N}$. Similarly, all terms involving fourth-order
derivatives cancel out. Consequently, each term in the l.h.s.~of \rf{diver}
gives rise to a single term
$\displaystyle\varepsilon_{i_1...\,i_N}\varepsilon_{j_1...\,j_N}
u_{x_{i_1}x_{j_1}}...\,u_{x_{i_N}x_{j_N}}$, and hence
the sum is the determinant of the Hessian of $u$, as required.

This form has an interesting consequence. Suppose, for the sake of simplicity,
that a space-periodic solution $u\in T^N$ is sought. Let $\varphi$ be a smooth
function with a finite support (in this context such $\varphi$ are called test
functions). Multiplying the MAE by $\varphi$ and {\em twice} integrating
by parts each term in the sum, we obtain by virtue of \rf{diver}
\begin{equation}
{1\over N!}\,\sum_{i_1,...,i_N,j_1,...,j_N}\varepsilon_{i_1...\,i_N}
\varepsilon_{j_1...\,j_N}
\int_{R^N}u_{x_{i_1}x_{j_1}}...\,u_{x_{i_{N-1}}x_{j_{N-1}}}u
\,\varphi_{x_{i_N}x_{j_N}}d{\bf x}=\int_{R^3}f\varphi\,d{\bf x}.
\label{intid}\end{equation}
As usual in the theory of partial differential equations,
a weak solution to \rf{MAeq}, $u$, can be defined as a function satisfying
the integral identity \rf{intid} for any test function $\varphi$.
(Other definitions of weak solutions to the MAE are natural: generalised
solutions obtained by geometric constructions \cite{P75} and the so-called
viscosity solutions \cite{Caf}; the two definitions are equivalent \cite{Gut}.)
Commonly (e.g., see \cite{Feng07a,DG06}), only {\em one} integration by parts is
performed in this integral identity. Our form is advantageous in that a lesser
regularity of the weak solution is required for \rf{intid} to be well-defined.
The following argument
illustrates this for $N>2$: For any $u$ from the Sobolev space $W^2_{N-1}(T^N)$
the integrals in the l.h.s.~of \rf{intid} are well-defined (because by
the Sobolev embedding theorem this implies $\nabla u\in L_{2(N-1)}(T^N)$ and
hence $u\in L_\infty(T^N)$\,).
By contrast, integrals in the similar identity obtained by just a single
integration by parts are {\em not} well-defined for $u\in W^2_{N-1}(T^N)$.
(Note that a viscosity solution for fully nonlinear second-order equations
is only required to be continuous \cite{Caf}.)

\section{The Fourier integral form of the MAE; positivity and bounds of kernels}

\subsection{Derivation}

In the Monge--Amp\`ere--Kantorovich approach to the cosmological reconstruction
problems \cite{Fr02,Br03}, the MAE arises for the potential
of the inverse Lagrangian map, in which the function $f$ is a ratio between
matter density at the current epoch and at the much earlier epoch of matter-radiation
decoupling (at the earlier epoch, the distribution is very close to uniform); thus, $f>0$.

Here, we consider a problem in which the spatial mean of $f$ is positive.
Note that for odd $N$, if $\la f\ra<0$, the change of the unknown function $u\to-u$
reverses the sign of $f$ and $\la f\ra$ admits the desirable sign
(here $\la\cdot\ra$ denotes space averaging:
\begin{equation}
\la f\ra\equiv\lim_{R\to\infty}{1\over|B_R|}\int_{B_R}f({\bf x})\,d{\bf x},
\label{lim}\end{equation}
and $|B_R|$ is the volume of the ball $B_R\subset R^N$ of radius $R$).
Suppose that $u$ and $f$ have the same spatial periodicity;
integration of \rf{diver} over a periodicity cell $T^N$ then yields
$$\int_{T^N}f\,d{\bf x}=0,$$
which is incompatible with $\la f\ra\ne0$. Thus we assume henceforth
\begin{equation}
u=c\left(\,{|{\bf x}|^2\over2}+u'\right),
\label{subst}\end{equation}
where $u'$ has the periodicity of $f$ and a zero spatial mean (this being
just a normalisation), and $c$ is a constant. Substitution of \rf{subst}
into \rf{MAeq} and integration over a periodicity cell yields
\begin{equation}
c^N=\la f\ra.
\label{ave}\end{equation}
To derive this, note that $u_{x_ix_j}=c(\delta_{ij}+u'_{x_ix_j})$, where
$\delta_{ij}$ is the Kronecker symbol, and hence the derivatives of $u'$ in the
l.h.s.~of \rf{MAeq} are present only in $\det\|u'_{x_ix_j}\|$ and a linear
combination of minors of smaller sizes of the Hessian of $u'$. By the algebraic
transformation presented in the previous section each such minor can be
converted into a sum of second derivatives of products of $u'$ with its second
derivatives. Thus the spatial mean of the l.h.s.~of \rf{MAeq} over any
periodicity cell, and hence the mean defined by \rf{lim}, does not involve
$u'$ and is equal to $c^N$.

In a more general formulation, we seek a solution \rf{subst} satisfying
\rf{ave} and $\la u'\ra=0$,
assuming that $f$ and $u'$ can be represented as Fourier integrals:
$$\nabla^2u'=\int_{R^N}\tph(\bom)e^{i\bom\cdot\bf x}\,d\bom,\quad
u'=\int_{R^N}\tu'(\bom)e^{i\bom\cdot\bf x}\,d\bom,\quad
\hbox{where \ }\tu'(\bom)=-\tph(\bom)/|\bom|^2,$$
$$f/c^N=\int_{R^N}\tf(\bom)e^{i\bom\cdot\bf x}\,d\bom.$$
(If the problem, defined by \rf{MAeq}, \rf{subst} and \rf{ave}, is considered
for space-periodic $f$ and $u'$, integrals in wave vectors in what follows are
replaced by the respective Fourier sums.)

Eq.~\rf{detone} is equivalent to
$$\det\|u'_{x_ix_j}\|={1\over N!}\,\int_{R^N}...\int_{R^N}{\det}^2
\left\|{\bf i}_{\bom^1},...,{\bf i}_{\bom^{N-1}},
{\bf i}_{\bom-\sum_{n=1}^{N-1}\bom^n}\right\|$$
$$\times\left(\,\prod_{n=1}^{N-1}\tph(\bom^n)\right)
\tph\!\left(\bom-\sum_{n=1}^{N-1}\bom^n\right)
e^{i\bom\cdot\bf x}\,d\bom^1...\,d\bom^{N-1}d\bom,$$
where $\bf i_a$ denotes a unit vector in the direction of $\bf a$. Our
immediate goal is to derive a similar expression for the terms in \rf{MAeq}
of lower orders in $u'$. The term of order $m$ is
$${(-1)^m\over N!}\sum_{i_1,...,i_N,j_1,...,j_N}\!\!
\varepsilon_{i_1...i_N}\varepsilon_{j_1...j_N}\sum_{|\sigma|=m}\!\!
\left(\,\prod_{n:~i_n,j_n\in\sigma}\int_{R^N}\tu'(\bom)\omega_{i_n}\omega_{j_n}
e^{i\bom\cdot\bf x}\,d\bom\right)\prod_{n:~i_n{~\rm or~}j_n\in\!\!\!/\sigma}
\!\!\delta_{i_nj_n}$$
(here the sum $\sum_{|\sigma|=m}$ is over all subsets
$\sigma\subset\{1,...,N\}$ of cardinality $m$)
$$={(-1)^m\over m!}\,\sum_{1\le p_1<...<p_{N-m}\le N}\,\int_{R^N}...\int_{R^N}
\!\left(\,\sum_{j_1,...,j_m}\varepsilon_{j_1...j_mp_1...p_{N-m}}
\prod_{n=1}^m\bom^n_{j_n}\right)^{\!\!2}$$
$$\times\left(\,\prod_{n=1}^m\tu'(\bom^n)\right)\exp\left(i
\sum_{n=1}^m\bom^n\cdot{\bf x}\right)\,d\bom^1...\,d\bom^m$$
$$=\int_{R^N}...\int_{R^N}A_m\left({\bf i}_{\bom^1},...,{\bf i}_{\bom^{m-1}},
{\bf i}_{\bom-\sum_{n=1}^{m-1}\bom^n}\right)$$
\begin{equation}
\times\left(\,\prod_{n=1}^{m-1}\tph(\bom^n)\right)\tph\!\left(\bom-\sum_{n=1}^{m-1}\bom^n\right)
e^{i\bom\cdot{\bf x}}\,d\bom^1...\,d\bom^{m-1}d\bom,
\label{mth}\end{equation}
where
\begin{equation}
A_m({\bf i}^1,...,{\bf i}^m)\equiv{1\over m!}\,
\sum_{1\le p_1<...<p_{N-m}\le N}\,M^2_{p_1...p_{N-m}}({\bf i}^1,...,{\bf i}^m)
\label{defAm}\end{equation}
is the sum of squares of all minors of rank $m$,
$$M_{p_1...p_{N-m}}({\bf i}^1,...,{\bf i}^m)\equiv\sum_{j_1,...,j_m}
\varepsilon_{j_1...j_mp_1...p_{N-m}}({\bf i}^1)_{j_1}...\,({\bf i}^m)_{j_m},$$
obtained by crossing out rows of numbers $p_1<...<p_{N-m}$ from the
$N\times m$ matrix
$${\cal M}_m\equiv\left\|{\bf i}^1,...,{\bf i}^m\right\|,$$
comprised of $m$ columnar vectors ${\bf i}^1,...,{\bf i}^m$.

Therefore, the problem \rf{MAeq} is reduced after the substitution \rf{subst}
to the system of integral equations, which we call the {\it Fourier integral
form} of the problem \rf{MAeq} and \rf{subst}:
$$\tph(\bom)+\sum_{m=2}^N\int_{R^N}...\int_{R^N}A_m\left({\bf i}_{\bom^1},...,
{\bf i}_{\bom^{m-1}},{\bf i}_{\bom-\sum_{n=1}^{m-1}\bom^n}\right)$$
\begin{equation}
\times\left(\,\prod_{n=1}^{m-1}\tph(\bom^n)\right)\tph\!\left(\bom
-\sum_{n=1}^{m-1}\bom^n\right)d\bom^1...\,d\bom^{m-1}=\tf(\bom),
\label{int}\end{equation}
which is now stated in terms of the Fourier coefficients $\tph(\bom)$
of the Laplacian of the unknown function $u'$. Equations \rf{int} are valid
for all $\bom\ne0$; the respective equation for $\bom=0$ is \rf{ave}.

If the r.h.s.~of \rf{MAeq} is zero-mean ($\la f\ra=0$), the MAE admits
the Fourier integral form \rf{int} (where $\tph(\bom)$ are now the Fourier
coefficients of $\nabla^2u$), where the l.h.s.~is reduced
to a single term for $m=N$ in the sum $\sum_{m=2}^N$.

\subsection{Bounds for the kernels $A_m$ in the Fourier integral form}

In this subsection we establish bounds
\begin{equation}
0\le A_m({\bf i}^1,...,{\bf i}^m)\le{1\over m!},
\label{ABbou}\end{equation}
provided all vectors ${\bf i}^s$ have a unit norm. These bounds will
play a crucial r\^ole for our numerical algorithm.

Addition to a column ${\bf i}^s$ of any linear combination of columns
${\bf i}^{s'}$ for $s'<s$ does not change the value of any minor
$M_{p_1...p_{N-m}}({\bf i}^1,...,{\bf i}^m)$. Using the Gram--Schmidt
orthogonalisation process, we change all ${\bf i}^s$ in ${\cal M}_m$ to
${\bf j}^s$ such that $(i)$ ${\bf j}^1={\bf i}^1$, $(ii)$ for any $s$,
${\bf j}^s$ differs from ${\bf i}^s$ by a linear combination of vectors
${\bf i}^{s'}$ for $s'<s$ and thus $A_m$ remains unaltered, $(iii)$ for any
$s$, ${\bf j}^s$ is orthogonal to all ${\bf j}^{s'}$ for $s'<s$. Hence
$${\bf j}^s={\bf i}_{{\bf j}^s}\sin\theta_s,$$
where $\theta_s$ is the angle between ${\bf i}^s$ and the subspace spanned by
$\{{\bf i}^{s'}|~s'<s\}$. Consequently,
$$A_m({\bf i}^1,...,{\bf i}^m)=
A_m(\,{\bf j}^1,...,{\bf j}^m)\prod_{s=2}^m\sin^2\theta_s.$$

We denote ${\cal M}'_m\equiv\left\|{\bf j}^1,...,{\bf j}^m\right\|$
and $^t{\cal M}'_m$ the transpose of ${\cal M}'_m$. The identity \cite{Kos}
\begin{equation}
\sum_{1\le p_1<...<p_{N-m}\le N}\,M^2_{p_1...p_{N-m}}({\bf j}^1,...,{\bf j}^m)
={\det}(^t{\cal M}'_m{\cal M}'_m)
\label{Kosdet}\end{equation}
(which does not require orthogonality of $\{{\bf j}^1,...,{\bf j}^m\}$)
can be easily proved using the formula
$$\det\|a_{ij}\|=\sum_{j_1,...,j_m}\varepsilon_{j_1...j_m}\prod_{i=1}^ma_{ij_i}.$$
We enlarge the set $\{{\bf j}^1,...,{\bf j}^m\}$ by vectors ${\bf j}^s$ for $s>m$
to a complete orthonormal basis in $R^N$. Let $\cal U$ be an orthogonal matrix
comprised of the $N$ columnar vectors ${\bf j}^1,...,{\bf j}^N$, and $\cal E$
be the $N\times m$ matrix, whose all entries are 0 except for ${\cal E}_{ss}=1$
for all $1\le s\le m$. Then ${\cal M}'_m=\cal UE$ and hence
$${\det}(^t{\cal M}'_m{\cal M}'_m)=\det(^t{\cal E}^t{\cal UUE})
=\det(^t{\cal E}{\cal E})=\det{\cal I}_m=1,$$
where ${\cal I}_m$ is the identity matrix of size $m$. Consequently,
$$A_m({\bf i}^1,...,{\bf i}^m)={1\over m!}\prod_{s=2}^m\sin^2\theta_s.$$
This demonstrates that \rf{ABbou} are sharp bounds.

\subsection{Solution of the MAE for a weakly fluctuating r.h.s.}

If the fluctuating part of the r.h.s.~in \rf{MAeq}, $f-\la f\ra$, is small
relative the mean $\la f\ra$, \rf{int} suggests that
$\tph(\bom)\approx\tf(\bom)$ and the nonlinear terms in \rf{int} are small.
Then the system can be solved by iteration:
$$\tph_{K+1}(\bom)=\tf(\bom)-\sum_{m=2}^N\int_{R^N}...\int_{R^N}A_m\left({\bf i}_{\bom^1},...,
{\bf i}_{\bom^{m-1}},{\bf i}_{\bom-\sum_{n=1}^{m-1}\bom^n}\right)$$
\begin{equation}
\times\left(\,\prod_{n=1}^{m-1}\tph_K(\bom^n)\right)\tph_K\!\!\left(\bom
-\sum_{n=1}^{m-1}\bom^n\right)d\bom^1...\,d\bom^{m-1}.
\label{linint}\end{equation}

\pagebreak
{\bf Theorem}

{\em $1^\circ$. Suppose positive constants $C_0$ and $C_1$ satisfy the inequality
\begin{equation}
\sum_{m=2}^N{(C_0+C_1)^m\over m!}\le C_1,
\label{cnst1}\end{equation}
\begin{equation}
\int_{R^N}|\tf(\bom)|\,d\bom\le C_0,
\label{bouphi}\end{equation}
and $\tph_0$ satisfies the inequality
\begin{equation}
\int_{R^N}|\tph_K(\bom)-\tf(\bom)|\,d\bom\le C_1
\label{bound}\end{equation}
for $K=0$. Then iterates \rf{linint} are globally bounded:
\rf{bound} holds true for all $K>0$.

$2^\circ$. Under the same conditions, the following inequalities are satisfied for all $K>0$:
\begin{equation}
\int_{R^N}|\tph_{K+1}(\bom)-\tph_K(\bom)|\,d\bom\le C_2
\int_{R^N}|\tph_K(\bom)-\tph_{K-1}(\bom)|\,d\bom,
\label{cont1}\end{equation}
\begin{equation}
\max_{\bom\in R^N}|\tph_{K+1}(\bom)-\tph_K(\bom)|\le C_2
\max_{\bom\in R^N}|\tph_K(\bom)-\tph_{K-1}(\bom)|,
\label{cont2}\end{equation}
where it is denoted
$$C_2\equiv\sum_{m=1}^{N-1}{(C_0+C_1)^m\over m!}.$$
Furthermore, if
\begin{equation}
C_2<1,
\label{cnst2}\end{equation}
then iterations \rf{linint} converge to a solution to \rf{int}, which is
unique in the ball \rf{bound}.}

The proof is elementary: Inequalities \rf{bound}--\rf{cont2} stem from
the bounds \rf{ABbou}. By the contraction mapping principle, inequalities
\rf{cont1}--\rf{cnst2} imply convergence of iterations \rf{linint} in the norms
of $L_1(R^N)$ and $C(R^N)$ in the space of Fourier coefficients,
yielding a solution satisfying \rf{bound}.

Evidently, equations \rf{cnst1} and \rf{cnst2}, where equality is assumed
instead of the inequalities, have positive solutions for any $N$. (For instance,
$C_0=\sqrt{3}-4/3,\ C_1=1/3$ for $N=3$.) If for the chosen values of $C_0$ and
$C_1$ \rf{cnst1} holds true, $C_2=1$ and the inequality \rf{bouphi} is strict,
then all conditions of the Theorem become satisfied for a slightly smaller
value of~$C_0$.

Note that in view of \rf{cnst2} $C_0<1$ and hence \rf{bouphi} implies $f>0$
everywhere. Consequently, our Theorem is mostly of interest as a statement
about convergence of iterations \rf{linint}. Existence of
weak solutions was proved by geometric methods in \cite{P75} for the MAE
with an arbitrary positive r.h.s.~in a compact convex domain. Although our
Theorem is significantly weaker because of the strong restrictions on
the r.h.s., we have proved it for a non-compact domain --- the entire space.

\section{The ``convolution'' form of the MAE}

Following the same algebraic ideas, the MAE can be partially ``integrated''.
We again use the Fourier transform of $u'$:
$$u'=\int_{R^N}\tu'(\bom)e^{i\bom\cdot\bf x}\,d\bom,\hbox{ \ where \ }
\tu'(\bom)=(2\pi)^{-N}\int_{R^N}u'({\bf x})e^{-i\bom\cdot\bf x}\,d\bf x.$$
Let
$$\txi(\bom)\equiv\sqrt{\tu(\bom)/(2\pi)^N},$$
where
\pagebreak
$$\arg(\txi(\bom))=\arg(\tu'(\bom))/2,\hbox{ \ if \ }|\arg(\tu'(\bom))|<\pi;$$
$$\arg(\txi(\bom))=-\arg(\txi(-\bom)),\hbox{ \ if \ }\arg(\tu'(\bom))=\pi.$$
That these conditions can be enforced, is elementary for $\bom\ne0$;
the case $\bom=0$ is not problematic, because $\txi(0)=0$. Then
$$\xi({\bf x})\equiv\int_{R^N}\txi(\bom)e^{i\bom\cdot\bf x}\,d\bom$$
is a real-valued function. (For a given $u'$ it is not uniquely defined.)

Let us render the term of order $m$ in $u'$ in the l.h.s.~of \rf{MAeq}
in the terms of $\xi({\bf x})$ employing the expression \rf{mth},
\rf{defAm}, \rf{Kosdet} and the ``identity''
$$(2\pi)^{-N}\int_{R^N}e^{i\bom\cdot\bf x}\,d\bom=\delta(\bf x)$$
(as usual understood in the sense of generalised functions):
$$\int_{R^N}...\int_{R^N}A_m\left({\bf i}_{\bom^1},...,{\bf i}_{\bom^{m-1}},
{\bf i}_{\bom-\sum_{n=1}^{m-1}\bom^n}\right)$$
$$\times\left(\,\prod_{n=1}^{m-1}\tph(\bom^n)\right)\tph\!\left(\bom-\sum_{n=1}^{m-1}\bom^n\right)
e^{i\bom\cdot{\bf x}}\,d\bom^1...\,d\bom^{m-1}d\bom$$
$$=(-1)^m\int_{R^N}...\int_{R^N}A_m\left(\bom^1,...,\bom^m\right)
\left(\,\prod_{n=1}^m\tu'(\bom^n)\right)
\exp\!\left(i\sum_{n=1}^m\bom^n\cdot{\bf x}\right)d\bom^1...\,d\bom^m$$
$$=\left(-(2\pi)^N\right)^m\int_{R^N}...\int_{R^N}A_m\left(
\txi(\bom^1)\bom^1,...,\txi(\bom^m)\bom^m\right)
\exp\!\left(i\sum_{n=1}^m\bom^n\cdot{\bf x}\right)d\bom^1...\,d\bom^m$$
$$=(2\pi)^{-Nm}\int_{R^N}...\int_{R^N}A_m\left(
\int_{R^N}\nabla\xi({\bf x})e^{-i\bom^1\cdot{\bf x}^1}\,d{\bf x}^1,...,
\int_{R^N}\nabla\xi({\bf x})e^{-i\bom^m\cdot{\bf x}^m}\,d{\bf x}^m\right)$$
$$\times\exp\!\left(i\sum_{n=1}^m\bom^n\cdot{\bf x}\right)d\bom^1...\,d\bom^m$$
$$={1\over m!}\int_{R^N}...\int_{R^N}\det\left(^t
\|\nabla\xi({\bf x}^1),...,\!\nabla\xi({\bf x}^m)\|
\|\nabla\xi({\bf x-x}^1),...,\!\nabla\xi({\bf x-x}^m)\|\right)
d{\bf x}^1...\,d{\bf x}^m.$$
In particular, the linear term ($m=1$) is
$\nabla^2u'=\int_{R^N}\nabla\xi({\bf x}^1)\cdot\nabla\xi({\bf x-x}^1)\,d{\bf x}^1$.

Therefore, the problem \rf{MAeq} is equivalent, after the substitution \rf{subst}
and \rf{ave}, to
\begin{equation}
1+\!\sum_{m=1}^N\!{1\over m!}\int_{R^N}\!\!\!...\!\!\int_{R^N}\!
\det\!\left(^t\|\nabla\xi({\bf x}^1),...,\!\nabla\xi({\bf x}^m)\|
\|\nabla\xi({\bf x\!-\!x}^1),...,\!\nabla\xi({\bf x\!-\!x}^m)\|\right)\!
d{\bf x}^1\!...\,d{\bf x}^m\!\!=\!\!{f\over c^N}.
\label{convo}\end{equation}
We call this integral equation the {\it ``convolution'' form} of the MAE. Since
it does not involve second-order derivatives, it may prove useful for development
of an iterative algorithm for numerical solution of the problem defined
by \rf{MAeq}, \rf{subst} and \rf{ave}, which involves suitable transformations
of the spatial variable.

If the r.h.s.~of \rf{MAeq} is zero-mean, the MAE also admits
the convolution form \rf{int}, where $c=1$, the l.h.s.~is reduced to a single
term for $m=N$ in the sum $\sum_{m=1}^N$, and $\tu'(\bom)$ in the definition
of $\txi(\bom)$ denotes the Fourier transform of $u$.

\section{A test problem with a cosmological flavour}

One important area of application of the MAE is the reconstruction of
the dynamical history of the Universe from present observations of the
spatial distribution of masses (galaxies, clusters, including their
dark-matter components). Let us briefly recall the background. Peebles
was the first to propose that, from the sole knowledge of the current
positions of galaxies (from the Local Group which includes our own
galaxy) without knowledge of their (proper) velocities, reconstruction
of the full dynamical history is a meaningful goal
\cite{Pe89}. Indeed, the very strong constraint on the distribution of
masses at the epoch of decoupling --- which has to be almost uniform
--- makes reconstruction a possibly well-posed two-point boundary
problem that can then be solved by variational techniques. In fact, using
convexity techniques, a theorem of uniqueness of reconstruction was
proved in \cite{Br03}, using the Euler--Poisson equations, which
describes the dynamics of matter on sufficiently large scales (of the
order of a few million light years). In practice, reconstruction on
such scales was done so far via the Monge--Amp\`ere--Kantorovich (MAK) method
\cite{Fr02} (see also \cite{Br03,Mo08}) which assumes that the Lagrangian map
from initial to current mass locations is the gradient of a convex
potential. This holds exactly for the Zel'dovich approximation
\cite{Zel} (in which the equation for the velocity of matter reduces
to the inviscid Burgers equation with a zero r.h.s., and hence the peculiar
velocities remain constant along trajectories of point masses)
and also for its refinement, the first-order Lagrangian
perturbation approximation \cite{MABPR91}. It is then easy, using mass
conservation, to derive a MAE for the potential of the inverse
Lagrangian map (the latter is the Legendre--Fenchel transform of the
potential of the direct map). By a theorem of Brenier
\cite{Br87} the MAE becomes a Monge--Kantorovich mass transportation
problem with quadratic cost which, after discretisation, can be solved
by optimisation techniques (see \cite{Br03} for details).
The r.h.s.~of the MAE is equal to the ratio of densities at the present and
initial positions. Thus, if the initial density was not uniformly distributed,
we would have to solve the MAE with the r.h.s.~depending on the gradient
of the unknown potential. However, at the epoch of decoupling (about 380 thousand years
after the Big Bang) the density of matter was close to uniform, which implies
a significant simplification of the MAE to be solved: its r.h.s.~is known.

Here we explore for the first time the possibility of directly solving
the three-dimensional MAE without discretisation on a toy model with a
cosmological flavour.

We assume that the dimension of space is $N=3$,
and the r.h.s.~of the MAE has the following structure:
\begin{equation}
f=\delta^{-3}\sum_{g=1}^Gf^{(g)}\left({\bf r-r}^{(g)}\over\delta\right).
\label{astro}\end{equation}
The function \rf{astro} describes mass distribution for $G$ ``objects'', which
are ``localised'', if the value of the parameter $\delta$ is small compared
to the distance between objects. In the cosmological context galaxies or clusters
of galaxies can be regarded as such objects; then $f^{(g)}$ describes the
{\it total} distribution of mass of both visible matter and dark matter,
in which the galaxies are embedded. Although observations attest, that
the distribution of visible matter in galaxies is clearly discontinuous,
it is astrophysically sound to expect that the distribution of all types
of matter is smooth due to the prevailing smoothness of the dark matter.
We assume that objects have density distributions with a Gaussian shape:
\begin{equation}
f^{(g)}({\bf r})={m^{(g)}\over(\sigma^{(g)}\sqrt{\pi})^3}\exp(-|{\bf r}/\sigma^{(g)}|^2).
\label{gau}\end{equation}
The $g$-th object of mass $m^{(g)}>0$ is located at ${\bf r}^{(g)}$. All
$m^{(g)}$, $\sigma^{(g)}$ and ${\bf r}^{(g)}$ are independent of the small
parameter $\delta>0$.

We will be seeking a solution \rf{subst} with a space-periodic $u'$
to the Monge--Amp\`ere problem \rf{MAeq}, where a space-periodic r.h.s.~$\wf$
is the sum of ``clones'' of \rf{astro} over all periodicity cells.
Without loss of generality we can assume that the total mass is normalised:
$$\int f({\bf r})\,d{\bf r}=\sum_{g=1}^Gm^{(g)}=1;$$
thus \rf{ave} implies $c=1$ in \rf{subst}.

It is instructive to consider particular solutions to this Monge--Amp\`ere
problem.

\subsection{An exact solution to the MAE for a spherically symmetric mass distribution}

In spherical coordinates centred at ${\bf r}^{(g)}$,
\rf{MAeq} becomes, for spherically symmetric $u$ and $f$,
$$\rho^{-2}{\partial^2 u\over\partial\rho^2}
\left({\partial u\over\partial\rho}\right)^2=\delta^{-3}f,$$
where $\rho=|{\bf r-r}^{(g)}|$. This equation has an obvious solution
\begin{equation}
u(\rho,\delta)=\int_0^\rho\left(3\int_0^{r'/\delta}r^2f(r)\,dr\right)^{1/3}dr',
\label{onestar}\end{equation}
whose first derivatives are uniformly bounded:
$${\partial u\over\partial x_i}=
{x_i\over\rho}\left(3\int_0^{\rho/\delta}r^2f(r)dr\right)^{1/3}=O(\delta^0);$$
the second derivatives are $O(\delta^{-1})$:
$${\partial^2 u\over\partial x_i\partial x_m}=\delta^{-1}{x_ix_m\over\rho^2}
\left({\rho\over\delta}\right)^2\!f\!\left({\rho\over\delta}\right)\!
\left(3\int_0^{\rho/\delta}\!r^2f(r)dr\right)^{-2/3}\!\!\!
+\left(\delta^i_m-{x_ix_m\over\rho^2}\right)\!{1\over\rho}\!
\left(3\int_0^{\rho/\delta}\!r^2f(r)dr\right)^{1/3}\!\!\!\!\!\!.$$
This estimate follows from the following inequalities:\\
$\bullet$ $|x_ix_m/\rho^2|\le 1$;\\
$\bullet$ for $\rho<\delta$, $f(\rho/\delta)$ is uniformly bounded and
$$\underline{c}(\rho/\delta)^3\le\int_0^{\rho/\delta}r^2f(r)dr\le
\overline{c}(\rho/\delta)^3$$
for some positive constant $\underline{c}$ and $\overline{c}$;\\
$\bullet$ for $\rho\ge\delta$, $(\rho/\delta)^2f(\rho/\delta)$ is uniformly bounded and
$$0<\underline{c}'\le\int_0^{\rho/\delta}r^2f(r)dr\le\overline{c}'$$
for some constant $\underline{c}'$ and $\overline{c}'$.

Moreover, if $\rho$ is larger than a positive constant,
$f(\rho/\delta)=o(\delta^3)$, and hence all second derivatives of $u$
are uniformly bounded outside any sphere of a fixed radius.

The properties of the solution discussed in this subsection are implied just
by a fast decay of $f$ at infinity and the spherical symmetry of the r.h.s.;
the assumption that the profile is Gaussian is unnecessary.

\subsection{A one-cell solution for $G>1$ spherically symmetric localised objects}

A solution for $G>1$ objects can be expected to admit a power series expansion
$$u({\bf r})=\sum_{n=0}u_n({\bf r},\delta)\delta^n,$$
where (by analogy with \rf{onestar}) each $u_n({\bf r},\delta)$ and its first
derivatives are $O(\delta^0)$, and the second derivatives are $O(\delta^{-1})$.

Naively one could expect interaction of localised objects to be asymptotically
unimportant, and hence to obtain in the leading order the sum of individual
one-object solutions:
\begin{equation}
u_0({\bf r})=\sum_{g=0}u^{(g)}(|{\bf r-r}^{(g)}|,\delta).
\label{zerosum}\end{equation}
Here $u^{(g)}(|{\bf r-r}^{(g)}|,\delta)$ is the one-object solution
\rf{onestar} for the $g$-th object. Let us inspect, to what extent
such a conjecture might be true.

Consider a neighbourhood of an object $\gamma$. In the leading order, the
l.h.s.~of \rf{MAeq} is $\det\|{u_0\,}_{x_ix_j}\|$. Substitution of \rf{zerosum}
into the determinant yields a sum of triple products of second derivatives
of various $u^{(g)}$. As shown in the previous subsection, for $g\ne\gamma$
each second derivative of $u^{(g)}$ is $O(\delta^0)$ in this neighbourhood
(the distance between the objects $g$ and $\gamma$ is a fixed positive
constant). Thus, only triple products of second derivatives of $u^{(\gamma)}$
contribute terms of the leading order, $O(\delta^{-3})$, the one of the l.h.s.
These products constitute $\det\|u^{(\gamma)}_{x_ix_j}\|$
and thus match the term $\delta^{-3}f^{(\gamma)}$ in the r.h.s.
Nonlinear interaction of pairs of solutions for individual objects is
at a lower order, $O(\delta^{-2})$, and that of triplets of one-object
solutions - still weaker, $O(\delta^{-1})$. Hence, after the finite
sum \rf{onestar} is substituted into the MAE, the highest order terms do
cancel out, and in this respect the naive conjecture is confirmed.

Nevertheless, the one-cell solution \rf{zerosum} does not represent the leading
order term of the solution to the problem at hand. To see this, note that at large
distances \rf{zerosum} exhibits a linear growth in $|\bf r|$, and not a quadratic
one, as the form of the solution \rf{subst} requires. Furthermore, after
subtraction of the quadratic profile $c|{\bf r}|^2$, the remaining part $u'$
must be space-periodic, which is clearly not the case for \rf{zerosum}.
We cannot use \rf{zerosum} to construct a global solution conformant with this
periodicity requirement by the procedure, used to obtain the periodic
r.h.s.~$\wf$ from the individual density profiles \rf{astro}, because the sum
of periodically distributed ``clones'' of one-cell solutions \rf{zerosum} is
infinite. Moreover, for such hypothetical space-periodic sum there would be
infinitely many pairwise interactions between objects yielding
products of the order $O(\delta^{-2})$; hence, even if the sum of such
products is finite, it will not necessarily remain $O(\delta^{-2})$.

We observe that the pairwise interaction does not become weaker when the
distance between the interacting objects grows. Thus, even in the high-contrast
limit the solution that we are seeking does not reduce to individual
interactions between objects; their interaction is essentially collective,
which makes it hard to predict the structure of the solution.

\section{Computation of a solution to the MAE with a space-periodic r.h.s.}

In this section we present two iterative algorithms for numerical solution
of the space-periodic Monge--Amp\`ere problem. One version (AICDM), employing
numerical improvement of convexity and discrepancy minimisation stabilising
the iterative process, is suitable for computation of solutions for everywhere
positive right-hand sides $f$ (see Subsection 6.4). Another one (ACPDM),
involving continuation in parameter and discrepancy minimisation, does not
require this condition to be satisfied (see Subsection 6.2). They both rely
on the basic algorithm for iterative solution of a fixed point problem for
the MAE (see Subsection 6.1). Test applications of the two algorithms are
considered in Subsections 6.3 and 6.4\,. We assume $\la f\ra\ne 0$, however,
reformulation of ACPDM for the case of a zero-mean space-periodic r.h.s.~is
straightforward.

\subsection{A basic solver (FPAR)}

If the amplitude of fluctuation of $f$ is small compared to the mean, --- more
precisely, if \rf{bouphi} is satisfied, --- the Theorem (see Subsection 3.3)
establishes existence of a solution to \rf{MAeq}, that is a perturbation of
$c|{\bf x}|^2/2$. If, in addition, \rf{cnst2} holds true, the Theorem
guarantees that iterations defined by \rf{linint} converge to the Laplacian of
the perturbation; this offers an algorithm for numerical solution of the MAE.
If either of the conditions \rf{bouphi} or \rf{cnst2} are violated (which is
the practically interesting case), iterations \rf{linint} do not necessarily
converge. Different algorithms are necessary, taking this into account.

Suppose the kernels $A_m$ in \rf{int} are frozen and take constant values
$a_m$, respectively. Then the new system of equations \rf{int} can be
easily solved, since it is the Fourier form of the polynomial equation
$$\sum_{m=0}^Na_m\eta^m=f/\la f\ra$$
in the physical space. This observation suggests the following algorithm.
We express \rf{int} in the terms of $\eta=\nabla^2u'$ as
\begin{equation}
\sum_{m=0}^Na_m\eta^m=f/\la f\ra+F(\eta),
\label{eqn}\end{equation}
where
$$F(\eta)\equiv1+\sum_{m=1}^Na_m\eta^m-\det\|\nabla^{-2}(\eta_{x_ix_j})+\delta_{ij}\|,$$
and implement iterations
\begin{equation}
\sum_{n=0}^Na_m\eta_K^m=f/\la f\ra+F(\eta_{K-1})
\label{itera}\end{equation}
in the physical space.

The applicability of this algorithm depends crucially on the choice of the
coefficients $a_m$. In view of the positivity and the bounds for the kernels
\rf{ABbou}, we impose
$$0\le a_m\le{1\over m!}\quad\mbox{for~~}m>0.$$
Furthermore, it seems practical to set
\begin{equation}
a_1=1,\qquad a_m={1\over2m!}\quad\mbox{for~~}m>1,
\label{choo}\end{equation}
because for $a_1=1$ the linear in $u'$ term is treated exactly,
\begin{equation}
F(\eta)=\!\sum_{m=2}^N\int_{R^N}\!...\int_{R^N}\left(A_m({\bf i}_{\bom^1},...,
{\bf i}_{\bom^m})-a_m\right)\left(\,\prod_{n=1}^m\tph(\bom^n)\right)
\exp\!\left(i\sum_{n=1}^m\bom^n\!\cdot{\bf x}\right)d\bom^1...\,d\bom^m
\label{Fagain}\end{equation}
and the median values \rf{choo} of $A_m$ for $m>1$ minimise the ranges of
the kernels
$$A_m({\bf i}_{\bom^1},...,{\bf i}_{\bom^m})-a_m$$
in \rf{Fagain}. Since for the coefficients \rf{choo}
$$|A_m({\bf i}_{\bom^1},...,{\bf i}_{\bom^m})-a_m|\le a_m,\quad m\ge2$$
the l.h.s.~of \rf{itera} ``captures'' the nonlinear behaviour of the l.h.s.~of
\rf{int}, the algorithm has chances to converge. However, for the extreme
values of $A_m$ the values of the kernels in \rf{Fagain} are as large as
the respective medians, and therefore convergence of iterations \rf{itera},
\rf{choo} is not guaranteed.

At least two strategies can be proposed for setting the value of $a_0$ (which
is a free parameter in the sense that \rf{int} is not required to be satisfied
for $\bom=0$):\\
$1^\circ$. At each iteration $a_0$ is tuned, so that $\la\eta\ra=0$ (which
holds true for $\eta=\nabla^2u'$).\\
$2^\circ$. We set $a_0=0$.\\
Note that although in the subspace of space-periodic functions the inverse
Laplacian is defined for zero-mean scalar fields only, $\la\eta_K\ra$
is not required to vanish, because the mean is removed by differentiation
in the Hessian before the inverse Laplacian is evaluated.

For any odd $N$, the equation \rf{itera} in $\eta_K$ with the coefficients
\rf{choo} has a unique root for any r.h.s. To check this, it is enough
to establish that the derivative $D_N(\eta)$ of the l.h.s.~of \rf{itera} is
positive for any $\eta$. For the choice of coefficients \rf{choo},
\begin{equation}
D_N(\eta)-D_N'(\eta)={1\over2}\left(1+{\eta^{N-1}\over(N-1)!}\right).
\label{derid}\end{equation}
At a minimum $D_N'(\eta)=0$ and hence \rf{derid} implies that at the minimum
$D_N>0$, proving monotonicity of the l.h.s.~of \rf{itera}.
It can be shown similarly, that for any even $N$ the number of roots is 0 or 2.
Thus, the algorithm is guaranteed to be applicable for odd $N$ only.

It can be proved, that the Theorem also applies for iterations \rf{itera}
(with the same constant $C_1$ bounding the same norm of solutions).

In an application to a test problem inspired by cosmology (see Subsection 6.3),
this algorithm, which we call FPAR ({\it Fixed Point Algorithm for the
Regular part of the MAE}) produces a sequence of iterations, initially
converging, but subsequently blowing up: linear instability sets in,
and hence the respective unstable mode must be removed.

\subsection{A more advanced solver (ACPDM)}

The behaviour of the basic algorithm FPAR suggests that it should be embedded
as an engine within a more advanced algorithm. In this subsection we present
such an advanced solver, ACPDM ({\it Algorithm with Continuation in a Parameter
and Discrepancy Minimisation}).

Consider a generalisation of \rf{eqn}:
\begin{equation}
Q(\eta;p)=0,
\label{eqn2}\end{equation}
where
$$Q(\eta;p)\equiv\sum_{m=0}^Na_m\eta^m-f/\la f\ra-pF(\eta).$$
Here the coefficients \rf{choo} are assumed, and the new parameter $p$ is
confined to the interval $[0,1]$. For $p=0$, \rf{eqn2} is just a set
of polynomial equations of degree $N$; for $p=1$ it reduces to \rf{eqn} which
is equivalent to \rf{MAeq}. Continuation in the parameter $p$ is implemented:
\rf{eqn2} is solved for a set of values $p_j$, increasing from 0 to 1, and
an initial approximation of the solution for $p=p_j$ is obtained by
polynomial extrapolation of solutions for all $p_{j'}<p_j$. (Numerical
extrapolation requires performing quadruple precision computations,
if the number of nodes exceeds roughly a dozen.) For any $p$, solution
of \rf{eqn2} involves iterations similar to \rf{itera}:
\begin{equation}
\sum_{m=0}^Na_m\eta_K^m=f/\la f\ra+pF(\eta_{K-1}).
\label{itera2}\end{equation}
The r.m.s. discrepancy
$$d(\eta)\equiv\sqrt{\left\la\phantom{_|}\!\!\right(Q(\eta;p)-\la Q(\eta;p)\ra
\left)\phantom{_|}\!\!^2\right\ra}$$
is used in the termination condition. Note that $\la Q(\eta;1)\ra=0$ and hence
$$d(\eta)\equiv\sqrt{\left\la\phantom{_|}\!\!\right(\det\|\nabla^{-2}
\eta_{x_ix_j}+\delta_{ij}\|-f/\la f\ra\left)\phantom{_|}\!\!^2\right\ra}$$
for $p=1$; in particular, $d(\eta)=0$ for $p=1$, if and only if for
$u'=\nabla^{-2}\eta$ and the normalisation \rf{ave} the field \rf{subst} is
a solution to the MAE.

Let $(\cdot,\cdot)$ denote a scalar product and $\|\cdot\|$ the induced norm of
a scalar field: $\|{\bf v}\|\equiv\sqrt{({\bf v,v})}$.
In the test runs reported in the next subsection, the scalar product
of the functional Lebesgue space $L^2(T^3)$ has been assumed:
\begin{equation}
({\bf u,v})=\int_{T^3}u({\bf x})v({\bf x})d{\bf x}.
\label{lebeg}\end{equation}
Other products (with different weight functions introduced
in the above integral) have been also considered (see Subsection 6.4).

If $\eta'$ and $\eta''$ are approximate solutions to the generalised MAE
\rf{eqn2}, then
\begin{equation}
Q(\eta';p)-Q(\eta'';p)=A(\eta'-\eta'')+O(\|\eta'-\eta''\|^2),
\label{linQ}\end{equation}
where $A$ is the linearisation of \rf{eqn2} around the solution. Consequently,
the concept of minimisation of the residual in Krylov spaces and approaches
for its realisation can be borrowed from solvers for linear problems (such as
the Generalised Conjugate Gradients Method \cite{Axe}). ACPDM involves sequences
of stabilised iterations described below, which exploit this concept. To make
such an iteration, the following data is required:
an approximate solution $\eta_K$, and two sets of $S$ scalar fields,
$v_s({\bf x})$ and $w_s({\bf x})$, $0\le s\le S$, where $0\le S\le S_{\max}$
(as well as some other quantities computed at the previous stabilised
iteration), and $S_{\max}$ is a parameter of the algorithm. It is supposed that
all $v_s$ are mutually orthogonal with respect to the scalar product
$(\cdot,\cdot)$, and
\begin{equation}
v_s=Aw_s+O(\|w_s\|^2)
\label{approxAu}\end{equation}
for small $w_s$.

A {\it sequence of stabilised iterations of ACPDM} is initialised using the
current approximation $\eta_0$ by computing $\eta_1=\eta'_1$ as a solution to
\rf{itera2} for $K=1$, and setting $S=0$ (i.e. the sets $w_s({\bf x})$ and
$v_s({\bf x})$ are empty). For $K>1$, a {\it stabilised iteration of ACPDM}
consists of the following steps:\\
$i$. At each grid point in the physical space solve equation \rf{itera2}:
$$\sum_{m=0}^Na_m(\eta'_K)^m=f/c^N+pF(\eta_K).$$
$ii$. Compute $Q(\eta'_K)$.\\
$iii$. Orthogonalise $v'\equiv Q(\eta'_K)-Q(\eta'_{K-1})$ to all $v_s$,
$1\le s\le S$, with respect to the scalar product $(\cdot,\cdot)$, and set
$$v_{S+1}=v'-\sum_{s=1}^S{(v',v_s)\over(v_s,v_s)}v_s,\qquad
w_{S+1}=\eta'_K-\eta'_{K-1}-\sum_{s=1}^S{(v',v_s)\over(v_s,v_s)}w_s.$$
$iv$. Compute
$$\eta_{K+1}=\eta'_K-\sum_{s=1}^{S+1}{(Q(\eta'_K),v_s)\over(v_s,v_s)}w_s.$$
$v$. If $S<S_{\max}$, increase $S$ by 1; otherwise (i.e., if $S=S_{\max}$),
discard $v_1$ and $w_1$, and decrease by 1 the indices $s$ of the remaining
$S_{\max}$ fields $v_s$ and $w_s$.\\
$vi$. Compute the r.h.s.~of \rf{itera2} for $\eta_{K+1}$ substituted in place
of $\eta_{K-1}$, the field $Q(\eta_{K+1})$ and $d(\eta_{K+1})$. If
$d(\eta_{K+1})$ is less than a given small threshold, then $\eta_{K+1}$ is the
desired approximate solution and computation for the present $p$ is finished.
If $\|Q(\eta_{K+1})\|>\|Q(\eta_K)\|$, then the current sequence is terminated.

A few comments are in order. Clearly, $v_{S+1}$ and $w_{S+1}$ obtained in step
$iii$, possess the required properties: $v_{S+1}$ is $(\cdot,\cdot)-$orthogonal
to all $v_s$ for $s\le S$, and \rf{approxAu} holds for $v_{S+1}$ and $w_{S+1}$
by virtue of \rf{linQ}. The coefficients in the sum computed in step
$iv$ minimise the discrepancy $\|Q(\eta'_K-\sum_{s=1}^Sq_sw_s)\|$, assuming all
$q_sw_s$ are small and hence their quadratic contributions are negligible.
The minimisation plays a dual r\^ole: on the one hand, it stabilises basic
iterations \rf{itera2}, removing the instability modes as soon as they become
substantial; on the other, it significantly increases the efficiency of the
algorithm (up to a factor 20 compared to other algorithms relying on iterations
\rf{itera2}). The assumption that nonlinear terms are small can be
incorrect; also, as a result of accumulation of the neglected nonlinear terms
after a number of steps, at some stage $v_s$ can cease to approximate $Aw_s$
accurately enough. If the inequality $\|Q(\eta_{K+1})\|>\|Q(\eta_K)\|$ is found
to hold true in step $vi$, we interpret this as an indication that the adverse
effect of nonlinearity has become significant, and then the algorithm breaks
the current sequence. In order to reduce the adverse influence of nonlinearity,
a small number $S_{\max}$ may be chosen; in our runs, $S_{\max}=5$.

Now we can assemble the algorithm from the building blocks, discussed above.
ACPDM performs continuation in the parameter $p$. For a given $p$, it starts
by carrying out basic iterations \rf{itera2}. As soon as convergence slows down
(we have used the condition $d^2(\eta_K)>d^2(\eta_{K-1})/2$\,), the algorithm
switches to perform a sequence of stabilised iterations. If the sequence
terminates in step $vi$ because nonlinear effects became significant, the
algorithm proceeds by performing basic iterations \rf{itera2}. Usually it takes
a small number of them for the instability to set in, and as soon as ACPDM
detects that the inequality $d^2(\eta_K)>\alpha d^2(\eta_{K-1})$ holds true,
it starts a new sequence. Here $\alpha$ is a parameter, which can slightly exceed 1
(in order to allow the transients to die off and the dominant instability modes
to set in, so that the latter could be efficiently removed in subsequent
stabilised iterations); in our test runs we have used $\alpha=1.05$~.

\subsection{Application to a test problem inspired by cosmology}

We have chosen to employ a problem of the kind discussed in Section 5
as a test-bed for our algorithm, because cosmological applications of the MAE
are probably the most important ones. The positiveness of the r.h.s.~(implying
convexity of solutions \cite{P75}) is not required for application of the algorithm.

We are seeking a solution \rf{subst} with a space-periodic $u'$ to the
Monge--Amp\`ere problem \rf{MAeq}, where a space-periodic r.h.s.~$\wf$ is
a sum of ``clones'' of \rf{astro} over all periodicity cells. Our poor man's
Universe involves $G=3$ objects in the periodicity cell $T^3([0,1]^3)$, centred
at the origin: $-1/2\le x_i\le 1/2$. They are described by the following
parameter values in \rf{gau}:
$$m^{(1)}:m^{(2)}:m^{(3)}={1\over18}:{1\over27}:{1\over32},$$
$$\delta=1,\qquad\sigma^{(1)}=\sigma^{(2)}={1\over6},\quad\sigma^{(3)}={1\over8},$$
\begin{equation}
{\bf r}^{(1)}={1\over 4}(-1,1,-1),
\quad{\bf r}^{(2)}={1\over 4}(1,-1,-1),\quad{\bf r}^{(3)}={1\over 4}(-1,-1,1).
\label{univ}\end{equation}
Noting that $\wf({\bf x})$ achieves its maximum at ${\bf r}^{(3)}$ and its minimum
at $\displaystyle{1\over4}(1,1,1)$, we can estimate the contrast number of the problem as
$${16\over4\,(12\exp(-18)+8\exp(-18)+16\exp(-32))}\approx 1.313\times10^7$$
(the factor 4 in the denominator accounts for the replicas of \rf{gau}, located
in the neighbour periodicity boxes at a distance $1/\sqrt{2}$).

Iterations defined by \rf{itera2} have been performed on a uniform grid
comprised of $64^3$ points in the periodicity cell $T^3([0,1]^3)$. We have
evaluated $\det\|u'_{x_ix_j}+\delta_{ij}\|$ by the pseudospectral method
with dealiasing (for $N=3$ this requires computation of the derivatives
$u'_{x_ix_j}$ in the physical space on the twice finer grid involving $128^3$
points). Apparently dealiasing does not play any important r\^ole in these
computations; this is consistent with the fast decay of the energy spectrum
of $\eta=\nabla^2u'$ by 11 orders of magnitude.

We have tested both strategies for choosing $a_0$ (see Subsection 6.1), as well
as a hybrid strategy, in which $a_0=0$, but after solving \rf{itera} we reset
$\eta_K:=\eta_K-\la\eta_K\ra$. It turns out that the hybrid strategy and
$1^\circ$ are very close in the number of iterations
necessary to obtain a solution to the MAE to the same accuracy $d(\eta_K)<10^{-10}$.
However, this implies that the hybrid strategy is several times more efficient
in the terms of CPU time, since it requires just one evaluation of solutions
to \rf{itera}, while evaluation of $a_0$ following strategy $1^\circ$ requires
several such evaluations. The strategy $2^\circ$ has proved to be the most efficient
one, both in the terms of CPU time and the number of required iterations.
(The same nodes $p_j$ were used in all the test runs.) In the remaining part of
the subsection we discuss convergence of the advanced method with $a_0=0$.

\begin{figure}[t]
\centerline{\psfig{file=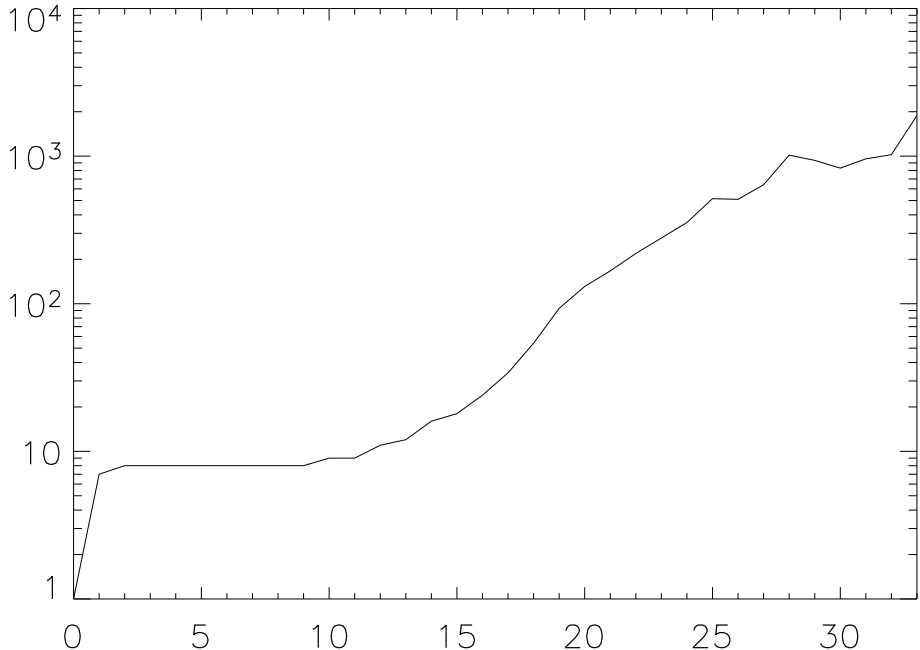,width=12cm}}

\caption{Number of evaluations of the determinant of the Hessian (vertical axis,
logarithmic scale) performed by ACPDM in successive computations of
approximate solutions $\eta(p_j)$ to the generalised MAE \protect{\rf{eqn2}},
satisfying $d(\eta(p_j))<10^{-10}$, for numerical
solution of the test MAE \protect{\rf{eqn2}} (see Subsection 6.3)
Horizontal axis: the index $j$ numbering consecutive nodes $p_j$ in the mesh
\protect{\rf{pj}}. The initial approximation for a $p_j$ is obtained
by the polynomial extrapolation of solutions for $p_{j'}$ with $j'<j$.}
\end{figure}

ACPDM performs a combination of basic iterations \rf{itera2} and stabilised
iterations, which have different computational costs. The most time-consuming
operation, determining the cost, is computation of the determinant of the Hessian
in $F(\eta)$. There are two such operations in a stabilised iteration, and
a single one in a basic iteration. Consequently, the former is approximately
twice longer than the latter. To enable an accurate comparison of
performance of various versions of our code, we report the number of evaluations
of determinant in each run (a run is terminated when the obtained approximate
solution $\eta$ satisfies the accuracy requirement $d(\eta)=10^{-10}$).
Note however, that the main bulk of computations is performed by doing
stabilised iterations; the number of basic iterations is usually below
1\% of the number of stabilised ones.

In all runs, the initial iteration is a solution to \rf{eqn2} for $p=0$.
If ACPDM is applied to this field for $p=1$, iterations quickly become chaotic
and cease to converge (this has necessitated to include into the algorithm
continuation in the parameter $p$).
Initially, ACPDM has been applied for uniformly distributed nodes $p_j=j/J$,
$j=0,...,J-1$ for $J=20$, for which the algorithm has shown a remarkably fast
convergence (see Fig.~1). A polynomial 20-node extrapolation yields
an approximate solution to our test problem to the accuracy
$d(\eta)=0.47\times10^{-2}$ and $d_\infty(\eta)=0.02$, where
$$d_\infty(\eta)\equiv\max_{T^3}\left|\det\|u'_{x_ix_j}+\delta_{ij}\|-\wf/\la\wf\ra\right|\!.$$
When it is used as an initial approximation for a run for $p=1$, convergence
is slow and the pattern of convergence is erratic. It takes 38\,685 evaluations
of the determinant of the Hessian for the two discrepancy norms
to decrease to $0.99\times10^{-9}$ and $0.39\times10^{-7}$, respectively.
At this stage convergence stalls. We have therefore got into a local
minimum of $d(\eta)$, out of which no exit can be found;
we will refer to it as a spurious minimum solution.

A better approximation to the solution \rf{subst} to our test problem
is obtained by adding more nodes near the right endpoint $p=1$.
We have chosen to add $J'=13$ nodes, the complete $p$-mesh being
\begin{equation}
p_j=j/J,\ \ j=0,...,J-1;\qquad p_{J+j-1}^{\phantom{2}}=1-(2^jJ)^{-1},\ \ j=1,...,J';\qquad
p_{J+J'}^{\phantom{2}}=1
\label{pj}\end{equation}
(information on convergence at the new nodes $p_{J+j}^{\phantom{2}}$ is also
included in Fig.~1). Initial approximations at each new $p_{J+j}^{\phantom{2}}$
become more and more accurate in the interval $0<p\le0.55$ ($j=11$), then
the discrepancy $d(\eta)$ of the initial approximations starts growing and
admits the maximum $0.52\times10^{-3}$ for $p=0.9875$ ($j=21$), and subsequently
decreases again. The rate of convergence progressively falls down as
$p_{J+j}^{\phantom{2}}$ approaches 1: the number of evaluations of
the determinant of the Hessian yielding solutions of the desired accuracy
$d(\eta(p_{J+j}^{\phantom{2}}))<10^{-10}$, markedly increases. However, ACPDM
does not stall any more. A polynomial extrapolation involving the 33 nodes
delivers an approximate solution $u'$ for $p=1$ to the accuracy
$d(\eta)=0.78\times10^{-7}$ and $d_\infty(\eta)=0.20\times10^{-5}$, and after
further 1\,885 evaluations of the determinant ACPDM yields an approximate
solution with the two discrepancy norms down to $10^{-10}$ and
$0.26\times10^{-8}$, respectively. In total, 9\,814 evaluations of
the determinant of the Hessian are involved in computations on the mesh \rf{pj}
providing the solution to the MAE.

The geometry of the obtained solution --- isosurfaces of $u'$ and $\nabla^2u'$ ---
is shown in Figs.~2 and 3, respectively. The structures disclosed by the
isosurfaces of $\nabla^2u'$ (Fig.~3) are clearly associated with the three objects
incorporated into the r.h.s.~\rf{astro} of the test problem. The figures also
reveal a subtle interaction of the objects along the lines connecting their
centres (see in Fig.~3 the tube-like structures, which connect the regions of
higher values of $\nabla^2u'$ encompassing the centres of objects).

\begin{figure}[p]
\centerline{\psfig{file=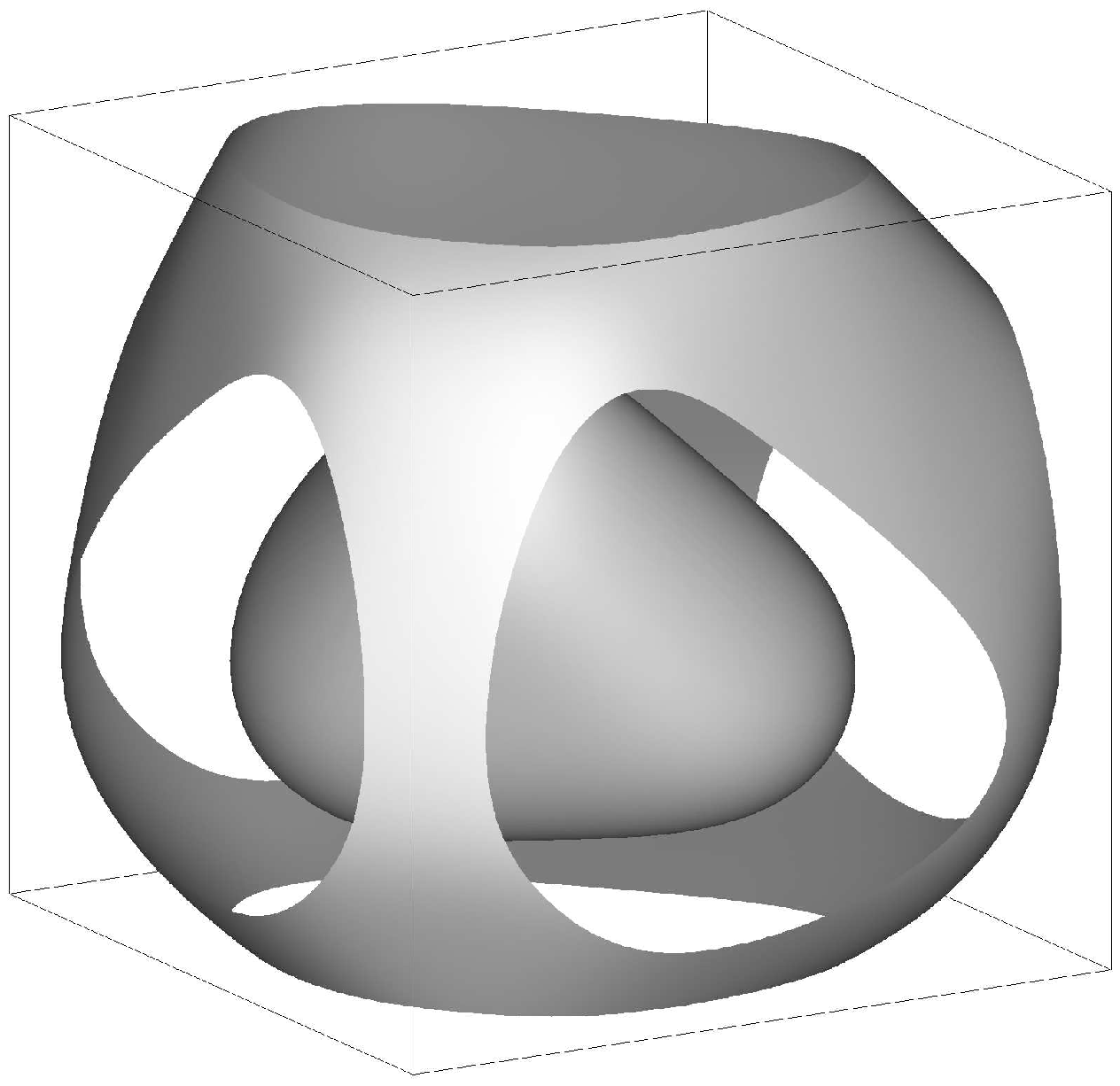,width=9cm,clip=}}
\centerline{(-0.5,-0.5,-0.5)\hspace*{25mm}}

\medskip
\caption{Isosurfaces of the solution \protect{\rf{subst}} to the test MAE
(presented in Subsection 6.3) at the levels of a half and
$1/8$ of the maximum. The periodicity cell $T^3([0,1]^3)$ of $u'$ is shown.}

\vspace{1cm}
\centerline{\psfig{file=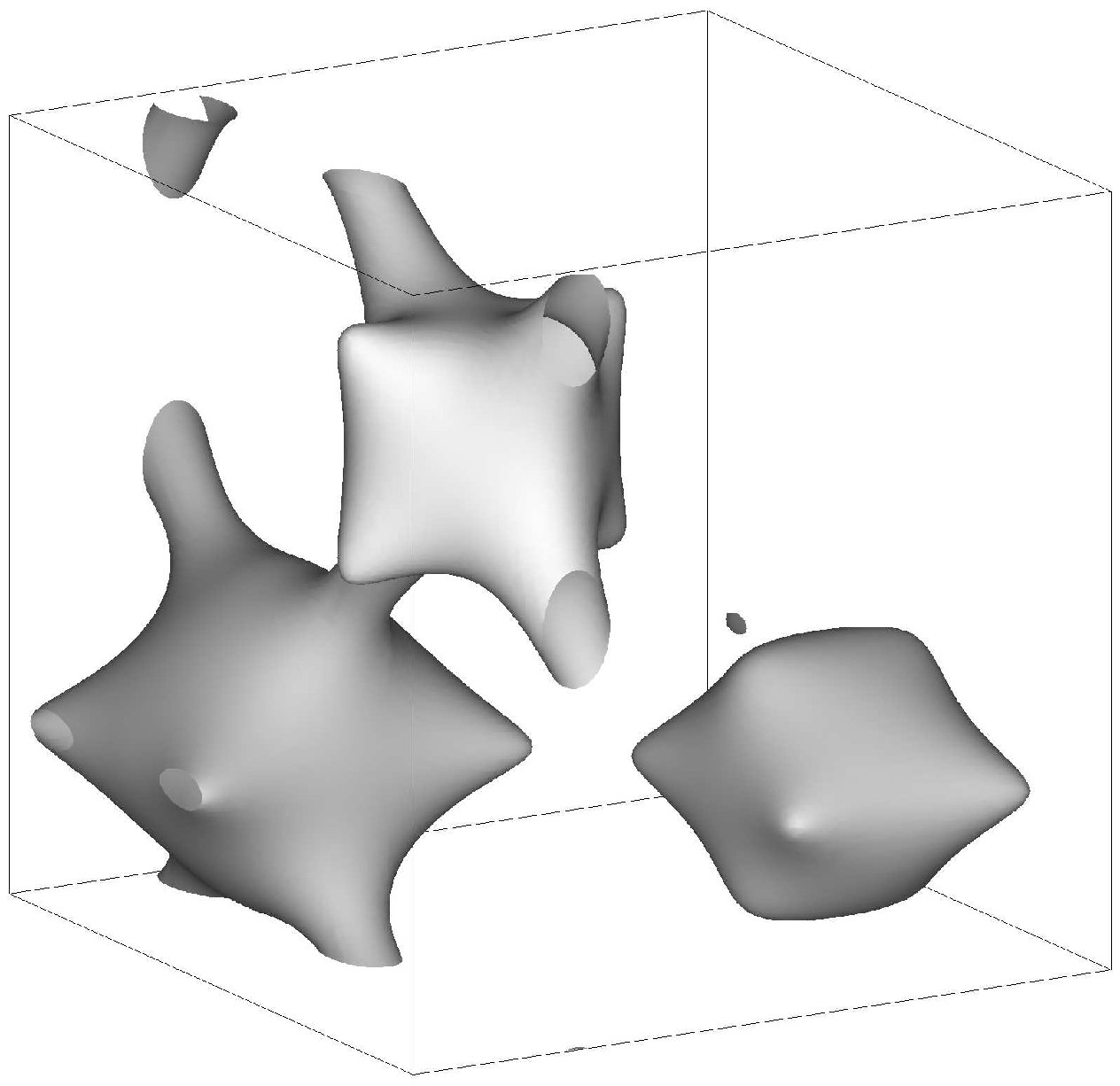,width=9cm,clip=}}
\centerline{(-0.5,-0.5,-0.5)\hspace*{25mm}}

\medskip
\caption{Isosurfaces of $\nabla^2u'$ for the solution to the test MAE at the
level of $1/3$ of the maximum. One periodicity cell $T^3([0,1]^3)$ is shown.}
\end{figure}

\begin{figure}[p]
\centerline{\psfig{file=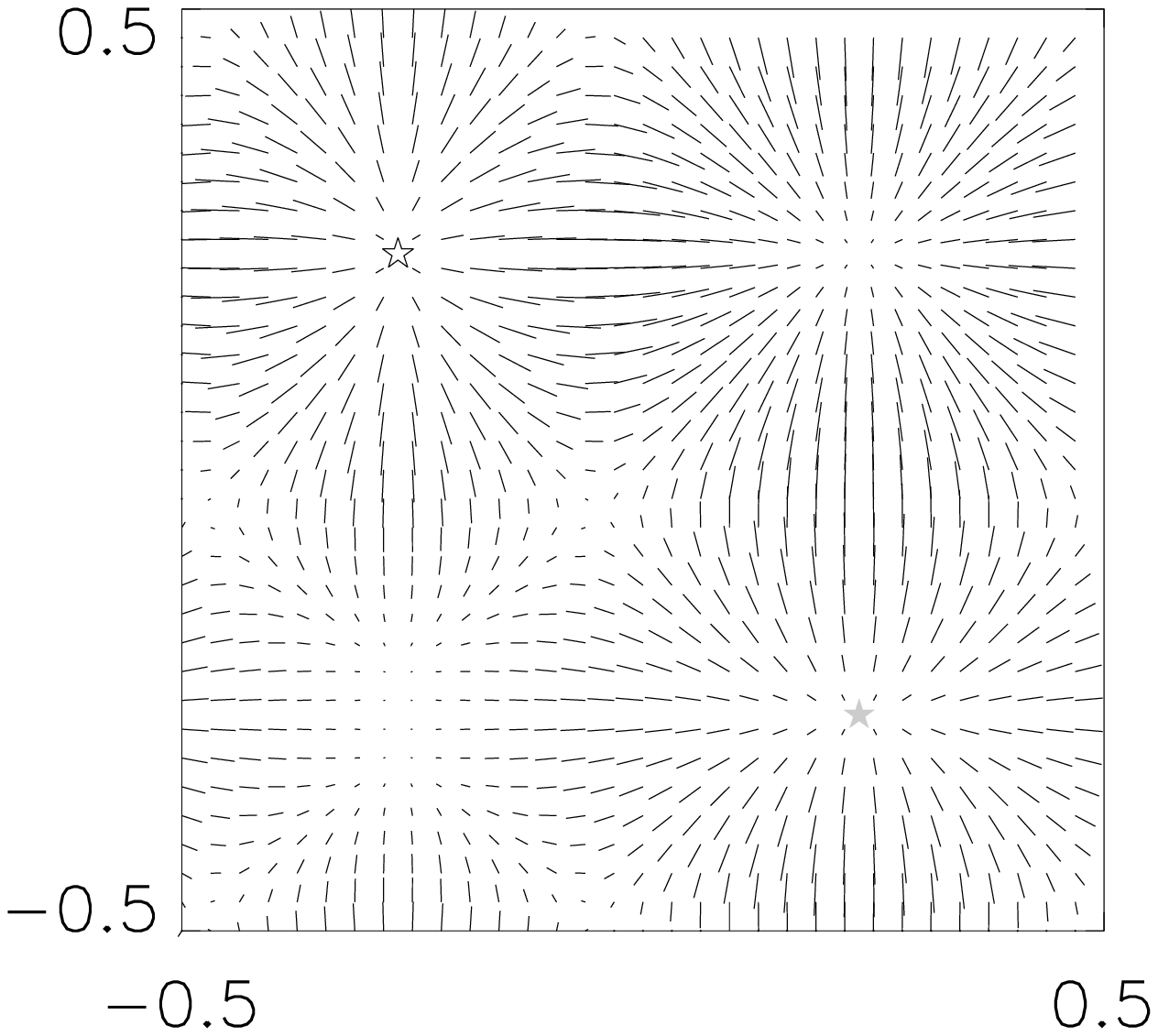,width=3.1in}\hfill
\psfig{file=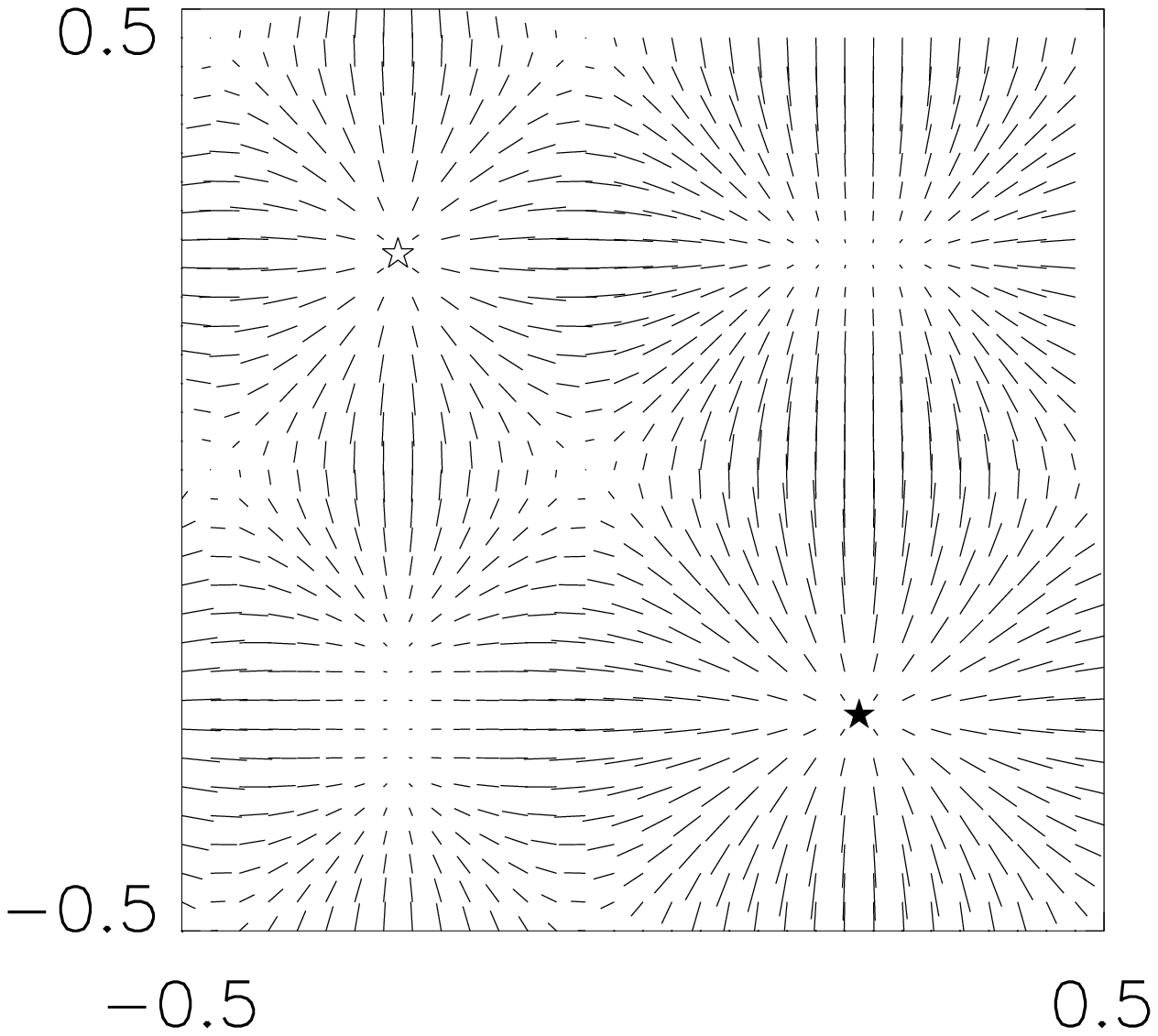,width=3.1in}}

\vspace*{-6mm}\hspace*{43mm}$x_1$\hspace*{77mm}$x_2$

\vspace*{-45mm}\hspace*{4.5mm}$x_3$\hspace*{77mm}$x_3$

\vspace*{44mm}
\centerline{(a)\hspace*{77mm}(b)}

\vspace*{1cm}
\centerline{\psfig{file=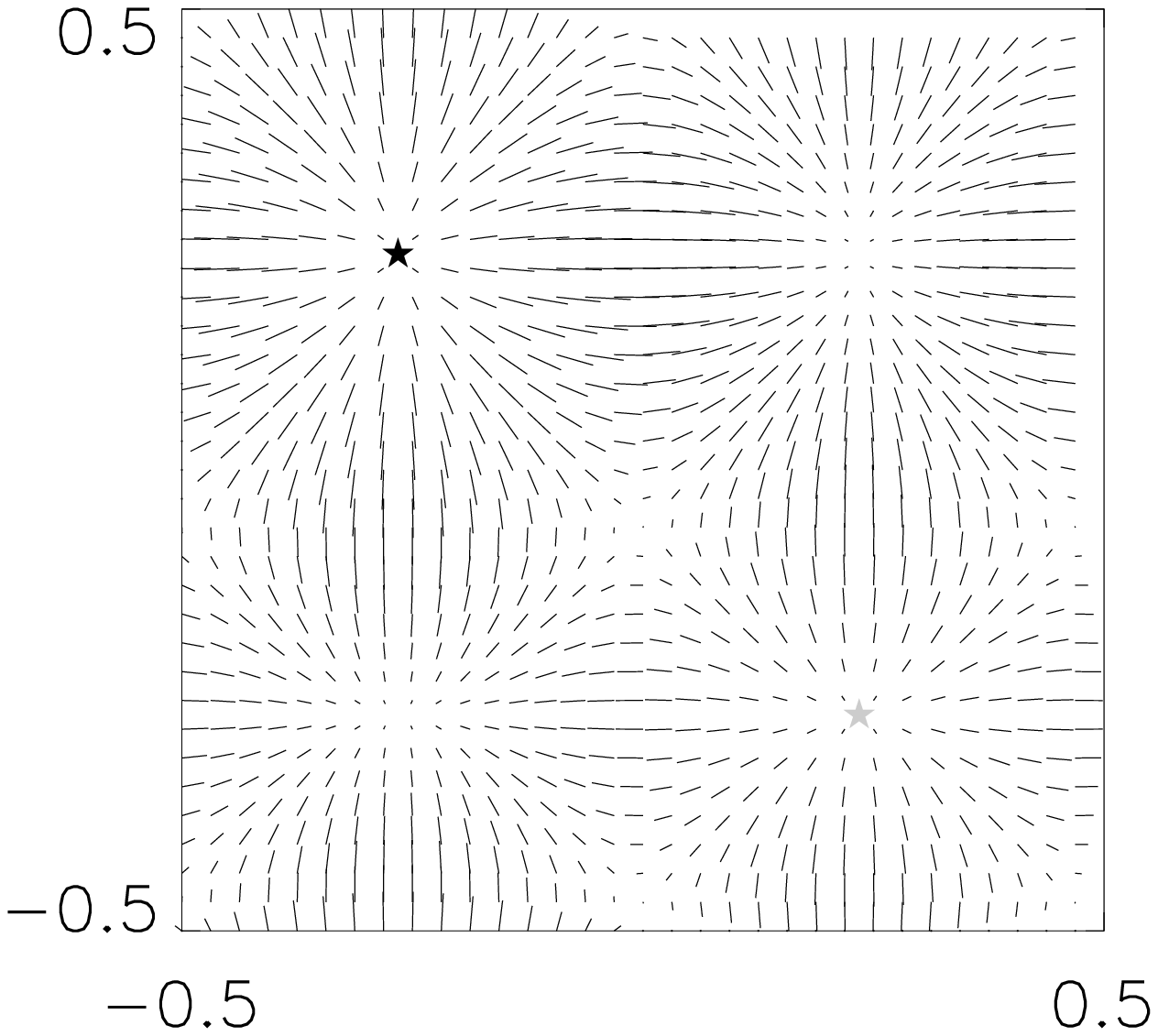,width=3.1in}}

\vspace*{-6mm}\hspace*{83mm}$x_1$

\vspace*{-45mm}\hspace*{45mm}$x_2$

\vspace*{44mm}\centerline{(c)}

\bigskip
\caption{$\nabla u'$ for the solution to the test MAE (presented
in Subsection 6.3) on cross sections of the periodicity cell $T^3([0,1]^3)$
that are parallel to coordinate planes and contain pairs of objects:
$x_2=-1/4$ (a), $x_1=-1/4$ (b), $x_3=-1/4$ (c). (Due to the symmetry of $u'$
about each of the three planes, components of gradients normal to the planes
are zero.) The labels $x_i$ refer to the Cartesian coordinate axes, parallel
to sides of the cross sections. Stars show locations of the three localised
objects \protect{\rf{univ}} on the cross sections. Gray-scaling reflects
the masses of the objects (black, gray and white stars: the objects
at ${\bf r}^{(g)}$, $g=1,2,3$, respectively).}
\end{figure}

\begin{figure}[p]
\centerline{\psfig{file=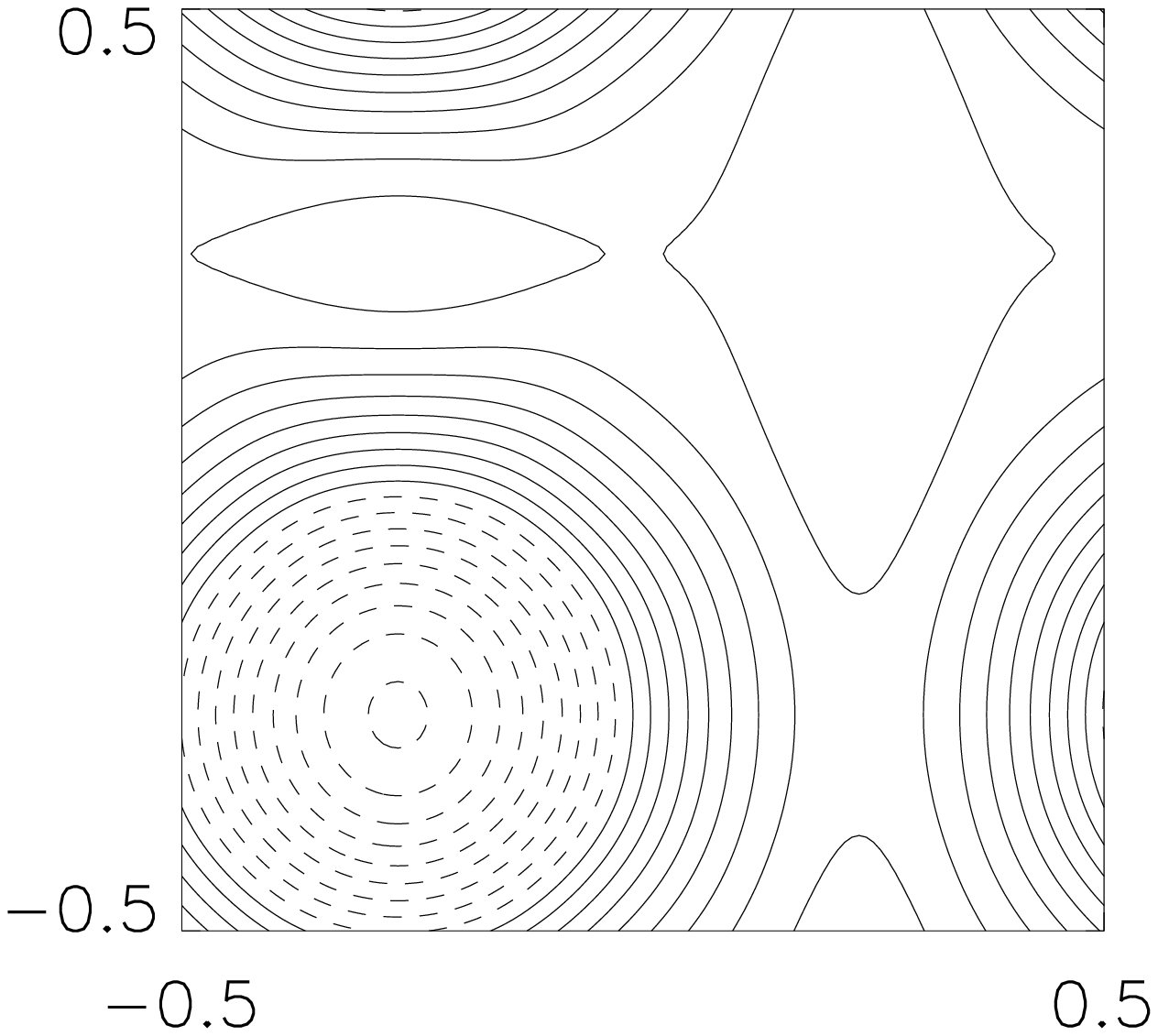,width=3.1in}\hfill
\psfig{file=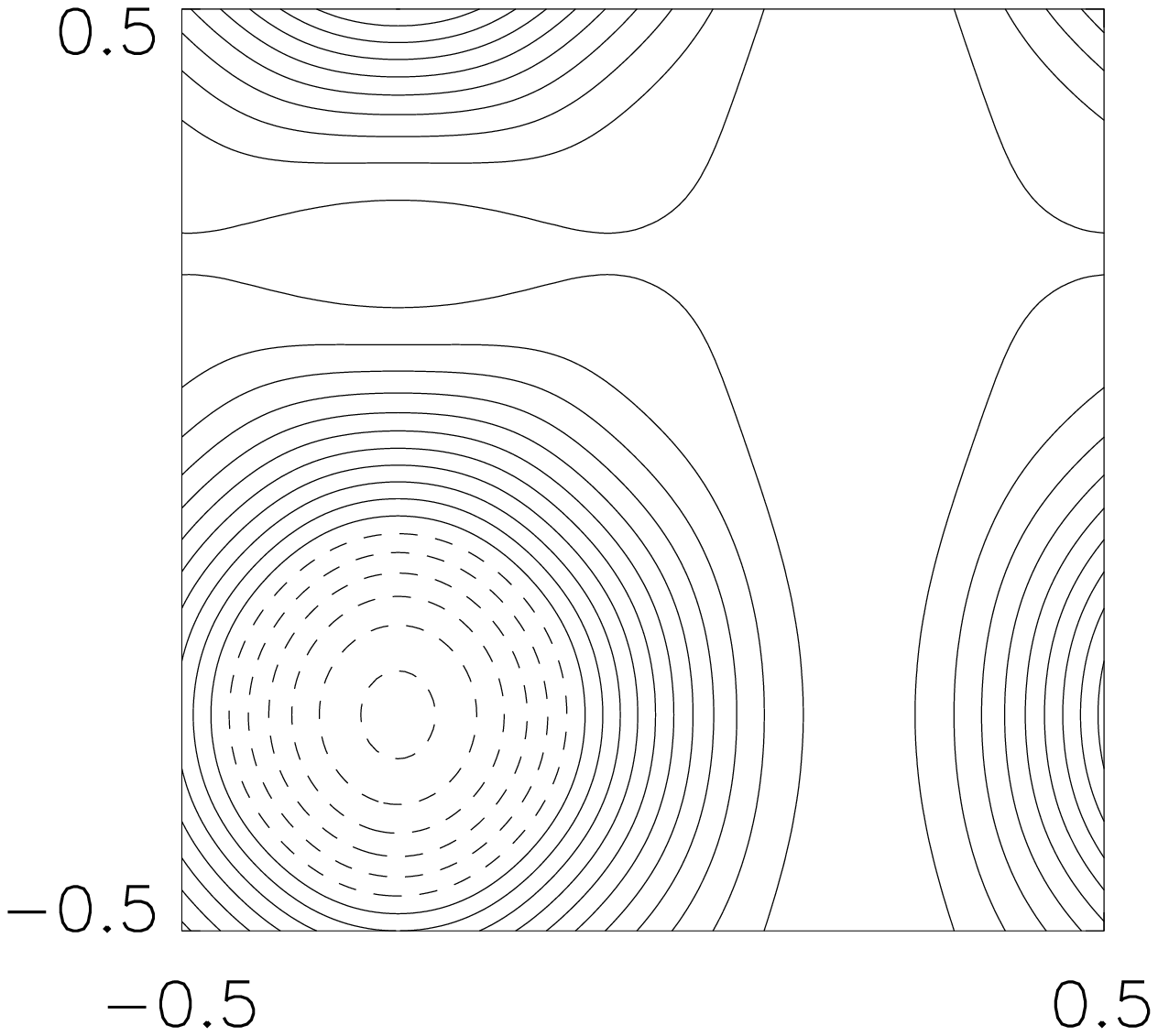,width=3.1in}}

\vspace*{-6mm}\hspace*{43mm}$x_1$\hspace*{77mm}$x_2$

\vspace*{-45mm}\hspace*{4.5mm}$x_3$\hspace*{77mm}$x_3$

\vspace*{44mm}
\centerline{(a)\hspace*{77mm}(b)}

\vspace*{1cm}
\centerline{\psfig{file=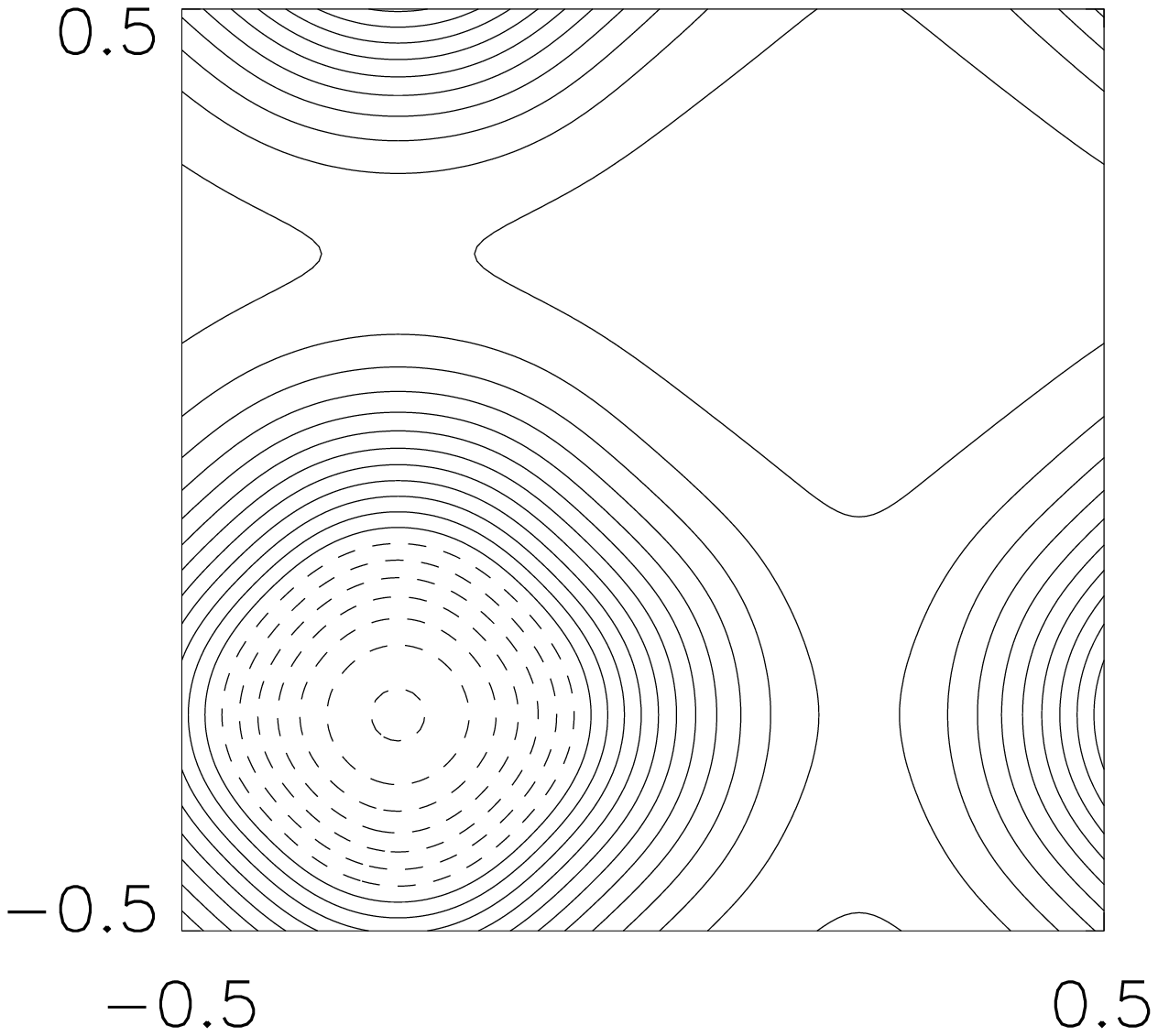,width=3.1in}}

\vspace*{-6mm}\hspace*{83mm}$x_1$

\vspace*{-45mm}\hspace*{45mm}$x_2$

\vspace*{44mm}\centerline{(c)}

\bigskip
\caption{Isolines step 0.02 of normal components of $\nabla u'$ for the solution
to the test MAE (dashed lines: negative values, solid lines: zero and positive
values) on Cartesian coordinate planes $x_2=0$ (a), $x_1=0$ (b), $x_3=0$ (c).
The labels $x_i$ refer to the Cartesian coordinate axes, parallel
to sides of the shown cross sections of the periodicity cell $T^3([0,1]^3)$.}
\end{figure}

Figures 4 and 5 display the structure of the field $\nabla u'$, determining
the displacement of mass in the test Universe.
It turns out that $u'$ possesses hidden symmetries, namely, mirror reflection
symmetries about any plane that is parallel to a coordinate plane and contains
a pair of objects. (Deliberately, our computations are not sped up by exploiting
these symmetries.) Because of the symmetries, the component of $\nabla u'$
normal to such a plane vanishes on this plane. To illustrate the behaviour
of the components of the gradient not shown in Fig.~4,
we present in Fig.~5 their isolines on coordinate planes, which pass through
the centre of the periodicity cube $T^3([0,1]^3)$ and hence are displaced
from the planes shown in Fig.~4 by a quarter of period.

\subsection{A solver for MAE with an everywhere positive r.h.s.~(AICDM)}

\tabcolsep1.2mm
\begin{table}[t]
\caption{Number of evaluations (NE) of the determinant of the Hessian,
performed by AICDM in the course of computation of approximate solutions
\protect{\rf{subst}} of varying accuracy to the test MAE with
the r.h.s.~defined by \protect{\rf{astro}}, \protect{\rf{gau}} and
\protect{\rf{univ}}, discrepancy $d_\infty(\eta)$ for these solutions, and
wallclock duration of runs (DR, seconds) for computation of approximate
solutions to the test MAE by EAICDM.}

\begin{center}
\begin{tabular}{|c|c|c|c|c|c|c|c|c|c|}\hline
$d(\eta)$\rule[0mm]{0mm}{4mm}&$10^{-3}$&$10^{-4}$&$10^{-5}$&$10^{-6}$&$10^{-7}$&$10^{-8}$&$10^{-9}$&$10^{-10}$\\\hline
$d_\infty(\eta)$\rule[0mm]{0mm}{4mm}&$0.64\!\times\!10^{-2}$&$0.86\!\times\!10^{-3}$&$1.02\!\times\!10^{-3}$&$0.23\!\times\!10^{-4}$
&$0.28\!\times\!10^{-5}$&$0.39\!\times\!10^{-6}$&$0.36\!\times\!10^{-7}$&$0.41\!\times\!10^{-8}$\\\hline
NE&64&126&251&439&1\,018&1\,515&2\,030&2\,653\\\hline
DR&3.38&4.69&11.17&42.7&81&203&435&774\\\hline
\end{tabular}\end{center}

\vspace*{6mm}
\caption{Number of evaluations (NE) of the determinant of the Hessian,
performed by AICDM with the use of various weight functions
\protect{\rf{wei}} in the scalar product $(\cdot,\cdot)$. All runs are terminated
as soon as the accuracy $d(\eta)<10^{-10}$ is obtained; $d(\eta)=0.66\times10^{-10}$
for $q=-1$, $d(\eta)=0.95\times10^{-10}$ for $q=-3/4$ and $d(\eta)=1.0\times10^{-10}$
in all remaining cases.}

\begin{center}
\begin{tabular}{|c|c|c|c|c|c|c|}\hline
$q$&$-1$&$-3/4$&$-1/2$&$-1/4$&0&1/4\\\hline
NE&3\,169&2\,619&2\,465&3\,743&2\,653&2\,571\\\hline
$d_\infty(\eta)$\rule[0mm]{0mm}{4mm}&$0.09\times10^{-8}$&$0.14\times10^{-8}$&$0.21\times10^{-8}$&$0.62\times10^{-8}$&$0.41\times10^{-8}$&$0.39\times10^{-8}$\\\hline
\end{tabular}

\vspace*{6mm}
\begin{tabular}{|c|c|c|c|c|c|c|c|c|}\hline
$q$&$1/2$&$3/4$&$1$&5/4&3/2&7/4&2\\\hline
NE&2\,568&2\,643&2\,523&2\,489&2\,481&2\,505&2\,526\\\hline
$d_\infty(\eta)$\rule[0mm]{0mm}{4mm}&$0.37\times10^{-8}$&$0.39\times10^{-8}$&$0.36\times10^{-8}$&$0.38\times10^{-8}$&$0.39\times10^{-8}$&$0.41\times10^{-8}$&$0.37\times10^{-8}$\\\hline
\end{tabular}\end{center}
\end{table}

Clearly, it is desirable to avoid the refinement of solutions for nodes
$0.95<p_j<1$, which has proved necessary in application of ACPDM to our test problem.
Comparison of the spurious minimum solution, obtained with the 20 nodes $p_j$,
and the more accurate one is instructive. The maximum discrepancy between the
two approximations of $\eta=\nabla^2u'$ for these solutions, equal to 0.132,
is attained at the minimum of the r.h.s., $(1/4,1/4,1/4)$, and all points, where
the discrepancy is larger than 0.02 are located within the distance 1/16 from
this point. The Hessian of \rf{subst} computed for the spurious minimum solution
is negative at this point, i.e. this solution \rf{subst} is not a convex
function, as it has to be \cite{P75}. This has suggested to develop a modification
of ACPDM, presented here, which we call AICDM ({\it Algorithm with Improvement of
Convexity and Discrepancy Minimisation}).

The algorithm proceeds as an extension of the ACPDM operating with a single
node $p=1$; computation of a good initial approximation by extrapolation in $p$
is unnecessary. AICDM involves an additional procedure: improvement of
convexity of an approximate solution, which is performed once the condition
$d^2(\eta_K)<\beta d^2(\eta_{K'})$
is satisfied. Here $K$ is the number of the current iteration, $K'$ is
the number of the iteration, at which the last previous improvement of
convexity was performed, and $\beta<1$ is a constant factor (we have chosen
$\beta=0.01$ in the test computation reported in this subsection).

Improvement of convexity in AICDM could be performed by computation of a convex
hull of the iterate, but we prefer to apply a numerically simpler procedure.
At each grid point in the physical space, we compute the eigenvalues $\lambda$
of the Hessian of $u'=\nabla^{-2}(\eta_K-\la\eta_K\ra)$ (they are all real,
since the Hessian is a symmetric matrix). If they are all larger than $-1$,
then the approximate solution given by \rf{subst} for the current iterate
$\eta_K$ is locally convex at the point under consideration (in this
discussion we assume that, after normalisation, $c=1$). Accordingly, if the
minimum eigenvalue $\lambda_{\min}$ exceeds $-1$, no action is taken. Suppose
now the contrary, i.e. $\lambda_{\min}<-1$. The minimum eigenvalue would become
$-1$, if at this point each second derivative $u'_{x_ix_i}$ is increased
by $-1-\lambda_{\min}$, and hence the Laplacian $\eta_K=\nabla^2u'$ is increased
by $3(-1-\lambda_{\min})$.
Following this observation, at the points where $\lambda_{\min}<-1$ we increase
$\eta_K$ by $6(-1-\lambda_{\min})$ (by choosing the factor 6 instead
of 3 we are ``overimproving'' $\eta_K$). It is not guaranteed, of course, that
$|{\bf r}|^2/2+\nabla^{-2}(\eta'_K-\la\eta'_K\ra)$ is convex for the resultant
$\eta'_K$, since this procedure changes all mixed derivatives at each point and
does not guarantee that each second derivative $u'_{x_ix_i}$ at the points of
local non-convexity is increased by the same amount, or that non-convexity
does not appear at new grid points. Hence we proceed, repeating the procedure
till each eigenvalue at each grid point becomes larger than $-1-d(\eta_K)/2$.
In our experience, only a small number of such iterations is necessary
to enforce this condition. The quantities $6(-1-\lambda_{\min})$ are small,
being at most comparable with the global discrepancy $d(\eta_K)$; although
the discrepancy $d(\eta'_K)$ for the ``improved'' approximation $\eta'_K$
can exceed the discrepancy $d(\eta_K)$ for the original approximation $\eta_K$,
it thus turns out that the increase (compared to $d(\eta_K)$) is usually modest.

An approximate solution to the test problem (formulated in Subsection 6.3),
satisfying $d(\eta)=10^{-10}$ and $d_\infty(\eta)=0.41\times10^{-8}$, has been
obtained by the AICDM. This has required 2\,653 evaluations of the determinant
of the Hessian. Table 1
illustrates the deterioration of convergence in this run, as better accuracy
numerical solutions are successively found. Note that computational cost
of each iteration in improvement of convexity of an approximate solution
slightly exceeds that of computation of the determinant of the Hessian, and each
iteration is included into evaluation counts (one evaluation per a convexity
improving iteration) presented in Tables 1 and 2.

In this run, as in runs reported in Subsection 6.3, the Lebesgue space scalar
product \rf{lebeg} has been assumed. We have also
inspected convergence of AICDM employing scalar products
$$({\bf u,v})=\int_{T^3}u({\bf x})v({\bf x})w({\bf x})d{\bf x}$$
with weight functions
\begin{equation}
w(x;q)=\max\left(1,({\wf/\la\wf\ra})^q\right)
\label{wei}\end{equation}
(see Table 2). For $q>0$, more prominence is given to discrepancy
in the regions, where the r.h.s.~of the MAE, $\wf$, admits relatively high
values; for $q<0$, the opposite happens, i.e. discrepancy in the regions, where
the values of $\wf$ are relatively low, is given more weight. Duration of
the shortest run, for $q=-1/2$, with a sequential code on a 3.16 GHz Intel Core
Duo processor is 25~min.~19~sec. The computations do not
reveal any clear dependence of the efficiency of AICDM on the power $q$ ---
in a large interval of $q$ the variation of the number of iterations is just
several per cent. The low sensitivity to the value of $q>0$ is probably linked
to the relative smallness of the region where $\wf>\la\wf\ra$.

\subsection{An enhanced version of AICDM}

The efficiency of AICDM can be further improved without employing new
mathematical ideas. For instance, we have explored the possibility of gradual
refinement of approximate solutions with increasing spatial resolution. The test
problem, considered in the two previous subsections, was solved
with the resolution of $(16M)^3$ Fourier harmonics, where the integer $M$ is
successively
increased from 1 to 4. When seeking a solution with the resolution $64^3$
harmonics, which satisfies the accuracy requirement $d(\eta)<10^{-\kappa}$,
the intermediate-stage computations with the resolution of $(16M)^3$ harmonics
are terminated as soon as AICDM finds a solution to the accuracy
$$d(\eta)<10^{-\min(\kappa,\,2.5M)}.$$
We call this algorithm enhanced AICDM, or EAICDM. Durations
of runs for integer $\kappa\ge3$ are shown in Table 2 (we do not present total
numbers of evaluations of the determinant of the Hessian in runs by EAICDM,
since the times of their computations vary significantly with the resolution).

A further acceleration is likely to be achieved by avoiding dealiasing,
but we did not explore this possibility.

\section{Concluding remarks}

We have presented new forms of the Monge--Amp\`ere equation in $R^N$:
the second-order divergence \rf{diver}, Fourier integral \rf{int} and
convolution \rf{convo} forms. They have been derived under the assumption that
the MAE \rf{MAeq} has a r.h.s.~with a non-vanishing spatial mean, and hence
\rf{MAeq} admits solutions \rf{subst}. The first form gives an opportunity
to relax the regularity requirements for weak solutions to the MAE, such that
the integral in the l.h.s.~of the identity \rf{intid} is well-defined.
The third form is an integro-differential equation of the first order, which
might be useful for development of an algorithm for computation of a solution
\rf{subst} using transformations of spatial variables. This paper
is mostly concerned with the Fourier integral form of the MAE used to prove
existence of a small-amplitude solution of the form \rf{subst} for a weakly
varying r.h.s.~in \rf{MAeq} and to formulate an algorithm for computation
of a solution \rf{subst} to \rf{MAeq} in an odd-dimensional space. In a test
application to a three-dimensional MAE with the r.h.s.~reminiscent of mass
transportation problems considered in cosmology, we have demonstrated that
a solution to the MAE with a smooth positive r.h.s.~can be efficiently obtained
by two versions of this algorithm, ACPDM (the algorithm with continuation
in a parameter and discrepancy minimisation) and AICDM (the algorithm
with improvement of convexity and discrepancy minimisation). While the latter
is suitable for computation of a convex solution for an everywhere positive
r.h.s.~$f$, the former requires only $\la f\ra\ne0$. However, modification of
ACPDM for the case of a zero-mean space-periodic r.h.s.~is straightforward.

There are some analogies between our method and the inexact Newton--Krylov
solver with preconditioning, used in \cite{Del08}. In our algorithms,
solving \rf{itera} or \rf{itera2}, where the easily invertible polynomial part
of the MAE is separated out, can be regarded
as preconditioning; we believe that it is optimal. Indeed, its design is based
on the algebraic nature of the MAE and primarily on the positivity and bounds
of the kernels in the Fourier integral form, a very special --- and so far
not reported --- property of the MAE. Furthermore, our
algorithms are of the Krylov type. However, we do not rely on Newton
iterations: instead of solving a succession of Newton problems, each
with a complexity comparable to that of the MAE, we tackle the MAE directly.

In this paper, space periodicity and Fourier decompositions play an important
r\^ole in two respects: First, we rely on the Fourier methods to find
the optimal coefficients \rf{choo} in the equations \rf{itera} or \rf{itera2}.
Note that space periodicity is not required for the derivations --- we
consider Fourier integrals, and not Fourier sums, and the algorithms do
not require computation of these integrals. Second, our algorithms are realised
in the spectral form, because we seek space-periodic solutions.
In a periodicity domain the required computations of the inverse Laplacian,
as well as numerical differentiation become trivial, when spectral methods are
applied. Recall that Fourier methods have also the advantage of being more
accurate: when the resolution is increased, Fourier series converge
to a solution together with all the derivatives that the solution possesses,
whilst finite differences do not approximate derivatives beyond their fixed
order. We would also use the spectral approach, for instance, to solve
numerically Dirichlet or Neumann boundary value problems for $u'$ (see
\rf{subst}) in a rectangular box.

However, the geometry of the domain, where the solution is sought, may prohibit
the use of spectral methods. What happens when the MAE must be solved for more
complicated boundary conditions in regions whose geometry is not rectangular
(parallelepiped-shaped)? Our algorithms
do not inherently rely on spectral methods and will remain applicable.
Since we are solving the MAE in terms of the Laplacian of the unknown function,
we would have to invert the Laplacian with suitable regular boundary conditions.
Finite differences can be applied in conjunction with our method to carry out
this task and for computation of derivatives involved in the function $F$
appearing in the r.h.s.~of
\rf{itera2} and \rf{itera2}. Key equations for our algorithms, \rf{itera} and
\rf{itera2}, are formulated in the physical space and can be used in conjunction
with our method for a variety of boundary conditions. Their convergence
properties remain of course to be investigated --- for instance,
whether the theoretically predicted values \rf{choo} will still be the optimal
choice for the coefficients of the polynomial in the l.h.s.~of \rf{itera2}.

An attractive feature of our algorithm is its simplicity: our Fortran-95
source code realising EAICDM is below 18 KB (672 lines long, not including
the source for the Fast Fourier Transform).

Hereafter we list some open questions. Under which conditions does
the generalised MAE
\begin{equation}
(1-p)\sum_{m=0}^Na_m\eta^m+p\det\|\nabla^{-2}\eta_{x_ix_j}+\delta_{ij}\|=f/\la f\ra
\label{gMAE}\end{equation}
with the coefficients \rf{choo}, which is solved for a set of $p$ by ACPDM,
possess zero-mean space-periodic solutions for all $0<p<1$? Does
its discretisation always have
a solution, as long as the generalised MAE itself does? Does the solution
of \rf{gMAE} depend analytically on the parameter $p$, and hence polynomial
extrapolation for $p=1$, that we employ, is mathematically sensible, or should
another asymptotics near $p=1$ be assumed? What is the optimal choice
of the sequence of values of $p$ to be used by ACPDM? Which scalar product is
optimal for acceleration of convergence of stabilised iterations? Can Chebyshev
techniques \cite{PZh} be used to improve efficiency of the iterative processes?

Additional questions emerging outside the main topic of this paper --- numerical
methods for solution of the MAE --- also cannot be avoided, since any information
concerning the structure of solutions to Monge--Amp\`ere problems
of the cosmological type with a high-contrast r.h.s., formulated in Section 5,
can be incorporated into specialised solvers in order to improve their
performance (like it has proved possible to accelerate computations about
4 times just by taking into account in AICDM convexity of solutions to the MAE
with a positive r.h.s.). The questions are: What is the asymptotics
of solutions in the small parameter $\delta$ determining the width
of the localised objects? Figures 4 and 5 show that the space is divided
into regions of dominant influence of each object. Also, notable are almost
axisymmetric structures in the Laplacian of the solution around the lines
connecting centres of objects, seen on Figure 3. Can these geometric features
be identified by performing the asymptotic analysis of the problem?
How does the contrast number measure numerical complexity
of the MAE and, in particular, the condition number of the linearisation
near the solution (for a given spatial discretisation)?

\medskip{\bf Acknowledgments}. We are grateful to S.~Colombi and A.~Sobolevskii for
discussions. Computations have been carried out on the computer ``M\'esocentre
SIGAMM (Simulations Interactives et Visualisation en G\'eophysique, Astronomie,
Math\'ematique et M\'ecanique)'' hosted by Observatoire de la C\^ote d'Azur,
France. Research visits of VZ and OP to the Observatoire de la C\^ote d'Azur
in the autumns of 2007 and 2008 were supported by the French Ministry
of Education. The authors were partially financed by the grant ANR-07-BLAN-0235
OTARIE from Agence nationale de la recherche (France). VZ and OP were also
partially supported by the grant 07-01-92217-CNRSL{\Large\_}a from the Russian
foundation for basic research.

\pagebreak


\begin{thebibliography}{99}
\bibitem{HTK}
S.~Haker, A.~Tannenbaum, R.~Kikinis. Mass preserving mappings and image registration.
Proc. of the 4th International Conference on Medical Image Computing and Computer-Assisted
Intervention. Lect. Notes in Comput. Sci., vol.~2208. Springer-Verlag, London (2001) 120--127.

\bibitem{HZTA}
S.~Haker, L.~Zhu, A.~Tannenbaum, S.~Angement. Optimal mass transport for registration
and warping. Int. J. Comput. Vision, 60 (2004) 225-240.

\bibitem{HGS}
T.~Hurtut, Y.~Gousseau, F.~Schmitt.
Adaptive image retrieval based on the spatial organization of colors.
Computer vision and image understanding, 112 (2008) 101--113.

\bibitem{Fr02}
U.~Frisch, S.~Matarrese, R.~Mohayaee, A.~Sobolevski. A reconstruction
of the initial conditions of the Universe by optimal mass transportation.
Nature, 417 (2002) 260--262 (arXiv:astro-ph/0109483).

\bibitem{Br03}
Y.~Brenier, U.~Frisch, M.~H\'enon, G.~Loeper, S.~Matarrese, R.~Mohayaee,
A.~Sobolevskii. Reconstruction of the early Universe as a convex optimization
problem. Mon. Not. R. Astron. Soc., 346 (2003) 501--524 (arXiv:astro-ph/0304214).

\bibitem{Mo08}
R.~Mohayaee, A.~Sobolevskii, The Monge--Amp\`ere--Kantorovich approach
to reconstruction in cosmology. Physica D, 237 (2008) 2145--2150 (arXiv:0712.2561).

\bibitem{GiTr83}
D.~Gilbarg, N.S.~Trudinger. Elliptic partial differential equations
of second order. Springer-Verlag, Berlin, 1983.

\bibitem{Bak94}
I.J.~Bakelman. Convex analysis and nonlinear geometric elliptic equations.
Springer-Verlag, 1994.

\bibitem{Caf}
L.A.~Caffarelli, X.~Cabr\'e. Fully nonlinear elliptic equations.
American Mathematical Society colloquium publications, vol.~43.
Amer. Math. Soc., Providence, Rhode Island, 1995.

\bibitem{OP88}
V.I.~Oliker, L.D.~Prussner. On the numerical solution of the equation
$\displaystyle{\partial^2z\over\partial x^2}\,{\partial^2z\over\partial y^2}
-\left({\partial^2z\over\partial x\partial y}\right)^2=f$
and its discretizations, I. Numerische Mathematik, 54 (1988) 271--293.

\bibitem{beam}
D.~Michaelis, S.~Kudaev, R.~Steinkopf, A.~Gebhardt, P.~Schreiber, A.~Br\"auer.
Incoherent beam shaping with freeform mirror. Nonimaging optics and efficient
illumination systems V. Eds. R.~Winston, R.J.~Koshel. Proc. of SPIE, vol.~7059
(2008) 705905.

\bibitem{GlOl}
T.~Glimm, V.~Oliker. Optical design of single reflector systems and the
Monge--Kantorovich mass transfer problem. J.~Math.~Sci. 117 (2003) 4096--4108.

\bibitem{P75}
A.V.~Pogorelov. The Minkowski multidimensional problem. Halsted Press,
Washington D.C., 1978. (Transl. from Russian: A.V. Pogorelov, The Minkowski
multidimensional problem. Nauka, Moscow, 1975.)

\bibitem{BB00}
J.-D.~Benamou, Y.~Brenier. A computational fluid mechanics solution to the
Monge--Kantorovich mass transfer problem. Numerische Mathematik,
84 (2000) 375--393.

\bibitem{DG03}
E.J.~Dean, R.~Glowinski. Numerical solution of the two-dimensional elliptic
Monge--Amp\`ere equation with Dirichlet boundary conditions: an augmented
Lagrangian approach. C. R. Acad. Sci. Paris, Ser. I, 336 (2003) 779--784.

\bibitem{DG04}
E.J.~Dean, R.~Glowinski. Numerical solution of the two-dimensional elliptic
Monge--Amp\`ere equation with Dirichlet boundary conditions: a least-squares
approach. C. R. Acad. Sci. Paris, Ser. I, 339 (2004) 887--892.

\bibitem{Feng07a}
X.~Feng, M.~Neilan. Galerkin methods for the fully nonlinear Monge--Amp\`ere
equation (arXiv:0712.1240).

\bibitem{Feng07b}
X.~Feng, M.~Neilan. Mixed finite element methods for the fully nonlinear
Monge--Amp\`ere equation based on the vanishing moment method (arXiv:0712.1241).

\bibitem{LR05}
G.~Loeper, F.~Rapetti. Numerical solution of the Monge--Amp\`ere equation
by a Newton's algorithm. C. R. Acad. Sci. Paris, Ser. I, 340 (2005) 319--324.

\bibitem{BFO09}
J.-D.~Benamou, B.D.~Froese, A.M.~Oberman. Two numerical methods
for the elliptic Monge--Amp\`ere equation. Preprint, 2009
[www.divbyzero.ca/froese/w/images/4/40/MA.pdf].

\bibitem{Del08}
G.L.~Delzanno, L.~Chac\'on, J.M.~Finn, Y.~Chung, G.~Lapenta. An optimal robust
equidistribution method for two-dimensional grid adaptation based
on Monge--Kantorovich optimization. J. Comput. Physics, 227 (2008) 9841--9864.

\bibitem{Finn}
J.M.~Finn, G.L.~Delzanno, L.~Chacon. Grid generation and adaptation
by Monge--Kantorovich optimization in two and three dimensions.
Proceedings of the 17th International Meshing Roundtable (2008) 551--568.

\bibitem{Gut}
C.E.~Guti\'errez. The Monge-Amp\`ere Equation. Progress in nonlinear
differential equations and their applications, vol.~44. Birkh\"auser, Boston,
2001.

\bibitem{Amp}
A.-M.~Amp\`ere. M\'emoire concernant ... l'int\'egration des \'equations
aux diff\'erentielles partielles du premier et du second ordre. Journal de
L'\'Ecole Royale Polytechnique 11 (1820) 1--188.

\bibitem{DG06}
E.J.~Dean, R.~Glowinski. Numerical methods for fully nonlinear elliptic
equations of the Monge--Amp\`ere type. Comput.~Methods Appl.~Mech.~Engrg.
195 (2006) 1344--1386.

\bibitem{Kos}
A.I.~Kostrikin. Introduction to algebra. Nauka, Moscow, 1977 (in Russian).

\bibitem{Pe89} P.J.E. Peebles.  Tracing galaxy orbits back in time.
Astrophys. J., 344 (1989) L53--L56.

\bibitem{Zel}
Ya.B.~Zel'dovich. Gravitational instability: an approximate theory for large
density perturbations. Astron. \& Astrophys. 5 (1970), 84--89.

\bibitem{MABPR91} F.~Moutarde, J.-M.~Alimi, F.R.~Bouchet, R.~Pellat,
A.~Ramani. Precollapse scale invariance in gravitational instability.
Astrophys. J. 382 (1991) 377--381.

\bibitem{Br87} Y.~Brenier. D\'ecomposition polaire et r\'earrangement
monotone des champs de vecteur. C.R. Acad. Sci.  Paris, Ser. I, 305 (1987) 805--808.

\bibitem{Axe}
O.~Axelsson. Iterative solution methods. Cambridge Univ. Press, 1996.

\bibitem{PZh}
O.M.~Podvigina, V.A.~Zheligovsky.
An optimized iterative method for numerical solution of large systems
of equations based on the extremal property of zeroes of Chebyshev polynomials.
J.~Scientific Computing, 12 (1997) 433--464.

\end{thebibliography}
\end{document}